\documentclass[sigconf, nonacm]{acmart}

\newcommand\vldbdoi{XX.XX/XXX.XX}

\newcommand\vldbavailabilityurl{https://github.com/intelligent-machine-learning/dlrover}
\newcommand\vldbpagestyle{plain} 

\renewcommand\footnotetextcopyrightpermission[1]{}
\AtBeginDocument{%
  \providecommand\BibTeX{{%
    \normalfont B\kern-0.5em{\scshape i\kern-0.25em b}\kern-0.8em\TeX}}}

\newcommand{\system}{{\sc DLRover-RM}\xspace}

\newcommand{\company}{{\sc AntGroup}\xspace}

\newcommand{\myline}[1]{{\medskip\noindent\textbf{#1}}}

\newcommand{\blacknumber}[1]{
    \tikz[baseline=(char.base)]{
        \node[shape=circle,fill=black,inner sep=0.7pt] (char) {\textcolor{white}{\bfseries #1}};
    }
}

\usepackage[linesnumbered,ruled,vlined]{algorithm2e} 
\usepackage{epstopdf}
\usepackage{multicol}
\usepackage{url}
\usepackage{enumerate}
\usepackage[labelfont=bf]{caption}
\usepackage{xcolor}
\usepackage{mathrsfs}
\usepackage{enumitem}
\usepackage{subfigure}
\usepackage{tabularx}
\usepackage{multirow}
\usepackage{diagbox}
\usepackage{pifont}
\usepackage{tikz}
\usepackage{amsmath} % for equation
\usepackage{hyperref}
\usepackage{graphicx} 
\usepackage{diagbox}
\usepackage[normalem]{ulem}

\useunder{\uline}{\ul}{}

\definecolor{dkgreen}{rgb}{0,0.6,0}
\definecolor{gray}{rgb}{0.5,0.5,0.5}
\definecolor{mauve}{rgb}{0.58,0,0.82}

\begin{document}

\title{\vspace{-3em}DLRover-RM: Resource Optimization for Deep
Recommendation Models Training in the Cloud}

%%
%% The "author" command and its associated commands are used to define the authors and their affiliations.

\thanks{* These authors contributed equally to this work. Jian Sha and Mingjie Tang are the corresponding authors. }

% \author{Qinlong Wang *}
% \affiliation{%
%   \institution{AntGroup}
% }
% \email{qinlong.wql@antgroup.com}

% \author{Tingfeng Lan *}
% \affiliation{%
%   \institution{Sichuan University}
% }
% \email{tafflan2001@gmail.com}

% \author{Yinghao Tang}
% \affiliation{%
%   \institution{Sichuan University}
% }
% \email{yinghaotang2001@gmail.com}

% \author{Bo Sang}
% \affiliation{%
%   \institution{AntGroup}
% }
% \email{b.sang@antgroup.com}

% \author{Ziling Huang}
% \affiliation{%
%   \institution{Sichuan University}
% }
% \email{youxingling1@gmail.com}

% \author{Yiheng Du}
% \affiliation{%
%   \institution{Sichuan University}
% }
% \email{yihengdu42@gmail.com}

% \author{Haitao Zhang}
% \affiliation{%
%   \institution{AntGroup}
% }
% \email{zht108229@antgroup.com}

% \author{Shajian}
% \affiliation{%
%   \institution{AntGroup}
% }
% \email{shajian@antgroup.com}

% \author{Hui Lu}
% \affiliation{%
%   \institution{The University of Texas at Arlington}
% }
% \email{hui.lu@uta.edu}

% \author{Ke Zhang}
% \affiliation{%
%   \institution{AntGroup}
% }
% \email{yingzi.zk@antgroup.com}

% \author{Mingjie Tang}
% \affiliation{%
%   \institution{Sichuan University}
% }
% \email{tangrock@gmail.com}
% \vspace{-3em}

% \author{
% Qinlong Wang$^{1,*}$, Tingfeng Lan$^{2,*}$, Yinghao Tang$^{2}$, Bo Sang$^1$, Ziling Huang$^2$, Yiheng Du$^2$, Haitao Zhang$^1$, Jian Sha$^1$, Hui Lu$^3$, Yuanchun Zhou$^4$, Ke Zhang$^1$, Mingjie Tang$^2$}

\author{
Qinlong Wang$^{1,*}$, Tingfeng Lan$^{2,*}$, Yinghao Tang$^{2}$, Ziling Huang$^2$, Yiheng Du$^2$, Haitao Zhang$^1$, Jian Sha$^1$, Hui Lu$^3$, Yuanchun Zhou$^4$, Ke Zhang$^1$, Mingjie Tang$^2$}

% \affiliation{%
%   $^1$Independent Researcher, \{qinlong.wql, zht108229, shajian, yingzi.zk\}@antgroup.com \\
%   $^2$Sichuan University, \{tafflan2001, yinghaotang2001, youxingling1, yihengdu42, tangrock\}@gmail.com \\
%   $^3$The University of Texas at Arlington, hui.lu@uta.edu, $^4$Chinese Academy of Science, zyc@cnic.cn.
% }

\affiliation{%
  $^1$Independent Researcher, 
  $^2$Sichuan University \\
  $^3$The University of Texas at Arlington,
  $^4$Computer Network Information Center, Chinese Academy of Science
}

%%
%% The abstract is a short summary of the work to be presented in the
%% article.

\label{sec:abstract}
\begin{abstract}
Deep learning recommendation models (DLRM) rely on large embedding tables to manage categorical sparse features. Expanding such embedding tables 
%the number of embedding rows 
can significantly enhance model performance, but at the cost of increased 
%training costs, such as high 
GPU/CPU/memory usage.
%especially in production environments. 
Meanwhile, tech companies have built extensive cloud-based services to accelerate training DLRM models at scale. In this paper, we conduct a deep investigation of the DLRM training platforms at \company and reveal two critical challenges: \textit{low resource utilization} due to suboptimal configurations by users and \textit{the tendency to encounter abnormalities} due to an unstable cloud environment. To overcome them, we introduce \system, an elastic training framework for DLRMs designed to increase resource utilization and handle the instability of a cloud environment.  
\system develops a resource-performance model by considering the unique characteristics of DLRMs and a three-stage heuristic strategy to automatically allocate and dynamically adjust resources for DLRM training jobs for higher resource utilization. Further, \system develops multiple mechanisms to ensure efficient and reliable execution of DLRM training jobs. 
%Additionally, we share best practices for training over 10,000 jobs in a production environment. 
Our extensive evaluation shows that \system reduces job completion times by 31\%, increases the job completion rate by 6\%, enhances CPU usage by 15\%, and improves memory utilization by 20\%, compared to state-of-the-art resource scheduling frameworks. \system has been widely deployed at \company and processes thousands of DLRM training jobs on a daily basis. \system is open-sourced and has been adopted by 10+ companies.

\end{abstract}

\maketitle

%%% do not modify the following VLDB block %%
%%% VLDB block start %%%
\pagestyle{\vldbpagestyle}
% \begingroup\small\noindent\raggedright\textbf{PVLDB Reference Format:}\\
% \vldbauthors. \vldbtitle. PVLDB, \vldbvolume(\vldbissue): \vldbpages, \vldbyear.\\
% \href{https://doi.org/\vldbdoi}{doi:\vldbdoi}
% \endgroup
% \begingroup
% \renewcommand\thefootnote{}\footnote{\noindent
% This work is licensed under the Creative Commons BY-NC-ND 4.0 International License. Visit \url{https://creativecommons.org/licenses/by-nc-nd/4.0/} to view a copy of this license. For any use beyond those covered by this license, obtain permission by emailing \href{mailto:info@vldb.org}{info@vldb.org}. Copyright is held by the owner/author(s). Publication rights licensed to the VLDB Endowment. \\
% \raggedright Proceedings of the VLDB Endowment, Vol. \vldbvolume, No. \vldbissue\ %
% ISSN 2150-8097. \\
% \href{https://doi.org/\vldbdoi}{doi:\vldbdoi} \\
% }\addtocounter{footnote}{-1}\endgroup
%%% VLDB block end %%%

% \vspace{-em}

%%% do not modify the following VLDB block %%
%%% VLDB block start %%%
\ifdefempty{\vldbavailabilityurl}{}{
\begingroup\small\noindent\raggedright\textbf{PVLDB Artifact Availability:}\\
The source code, data, and/or other artifacts have been made available at \url{\vldbavailabilityurl}.
\endgroup
}
%%% VLDB block end %%%

\section{Introduction}
\label{sec:introduction}

Deep learning based recommendation models (DLRM) are prevalent in recommendation scenarios~\cite{review_recommendation_model_1,review_recommendation_model_2,review_recommendation_model_3,deep_interest_network,deepfm,xdeepfm}. For example, Meta uses DLRMs for advertisement recommendation to optimize ad content for individual users, aiming to maximize click-through rates and advertising revenue~\cite{dlrm}. The training of DLRMs at Meta, Amazon, Alibaba, and \company can account for over 50\% of the total AI training cycles in cloud data centers \cite{acun2021understanding,facebook-dnn-rec,amazon-rec,ai-frontier}.

A typical DLRM uses {\it embedding tables} to manage sparse categorical features (e.g., User IDs) and several {\it deep neural networks} (DNNs) to improve the generalization of the models (\S\ref{subsec:DLRM_training_at_antgroup}). As the accuracy of a DLRM often improves with larger embeddings, which incorporate more feature data points, the size of DLRM embeddings has been steadily expanding, reaching up to terabytes with billions of embedding vectors \cite{acun2021understanding,lui2021understanding,zhou2019deep}. Tech companies build extensive cloud-based services to accelerate training these models at scale, e.g., with thousands of computing nodes \cite{review_cloud_computing_1,Morphling,mlaas}. Unfortunately, we observed from our cloud-based cluster that the resource utilization of over 80\% DLRM training jobs is under 50\%, indicating a significant \textit{underutilization} and waste of computation resources. Moreover, we observed that \textit{high instability} in cloud environments leads DLRM training to: 1) experience a high failure rate and 2) frequently encounter abnormalities  (e.g., stragglers) (\S \ref{subsec:challenges_of_DLRM_training_at_antgroup}).

In this paper, we focus on developing a highly resource-efficient and reliable {\it DLRM training system}, especially in a cloud environment, where failures are common and resource availability varies dynamically. Such a training system should be capable of training a multitude of DLRMs (e.g., 1,000s) concurrently with the following key goals: maximizing resource utilization (e.g., CPU and memory), achieving rapid training speeds, and ensuring fault tolerance. Achieving these stringent goals requires the training system to accurately {\it allocate} computational resources to individual DLRM training jobs and {\it schedule} these jobs in the cloud elastically and robustly. However, the unique characteristics of DLRMs, in combination with the dynamic nature of the cloud, make resource allocation and scheduling for DLRM training jobs extremely challenging.

%aiming for enhanced resource utilization (CPU and Memory) for model training and less job completion time (JCT). Given a DLRM job, we hope to predict its resource demand based on the model's characteristics and allocate the appropriate resources accordingly. 

\begin{figure}[h]
    \vspace{-2em}
    \hspace{-1.7em}
      \subfigure[CPU time distribution.] {
        \includegraphics[scale=0.125]{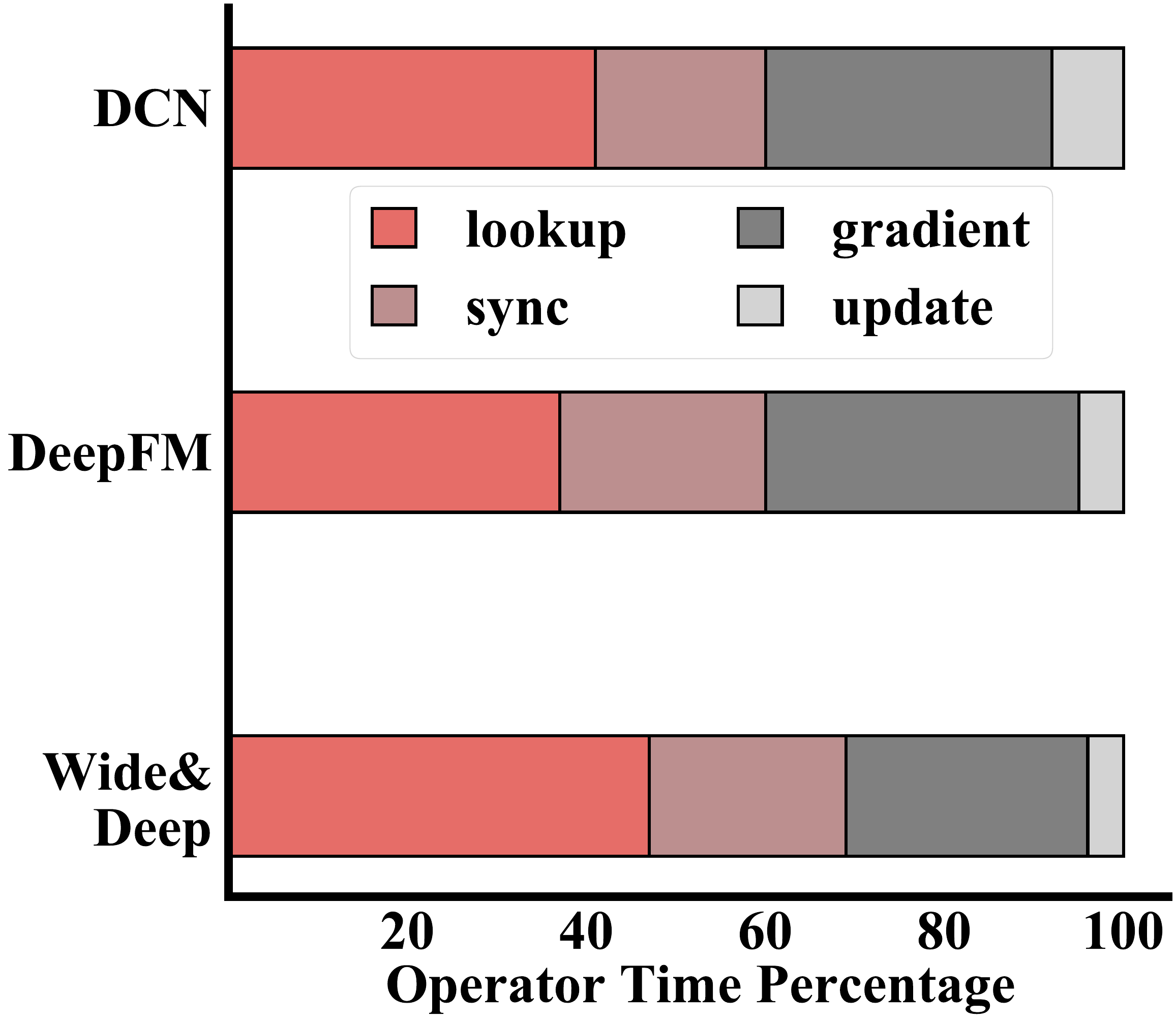}
        \label{fig:op_percentage}
      }
    \hspace{-2em}
    \subfigure[Memory size over time.] {
        \includegraphics[scale=0.125]{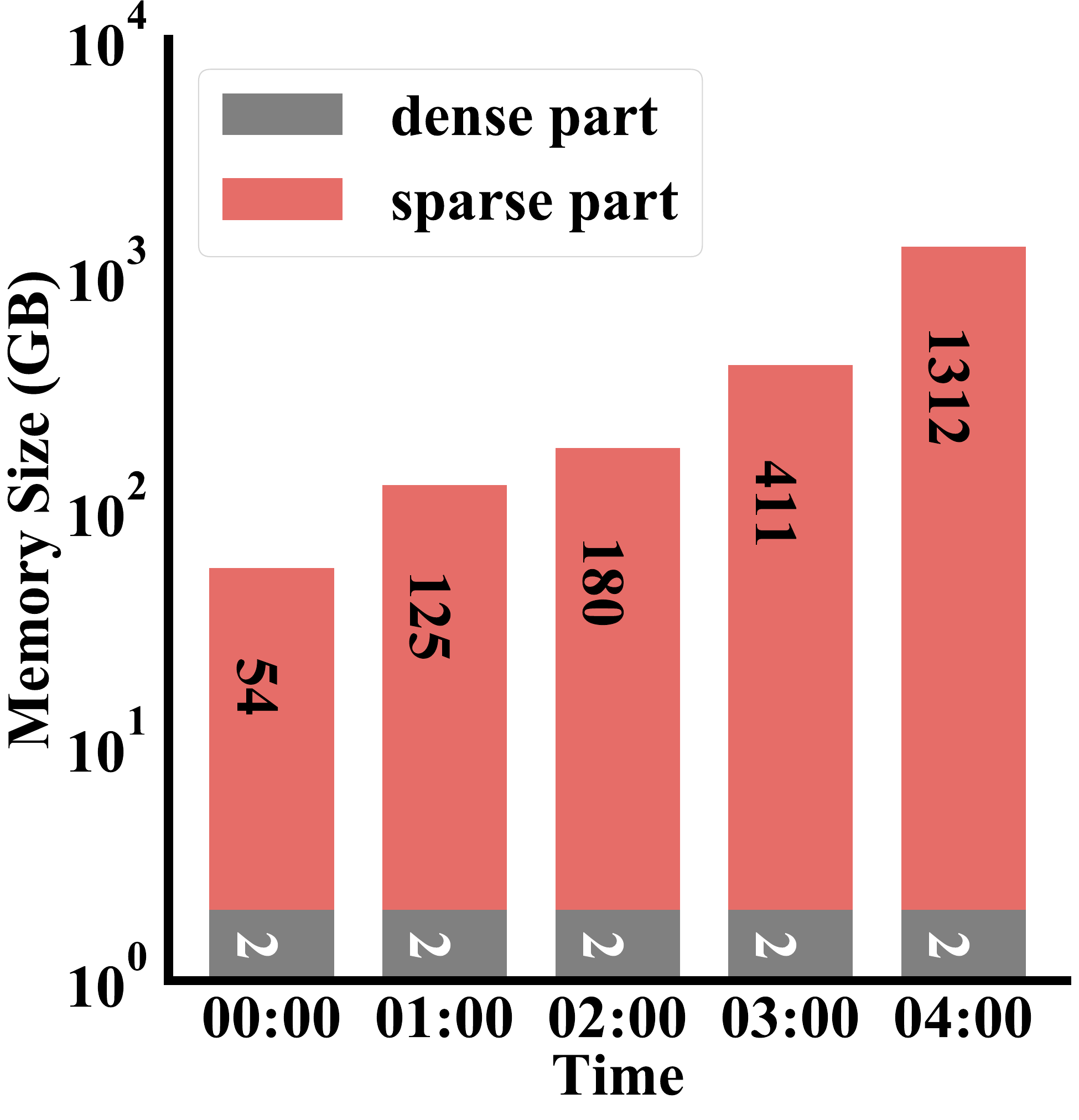}
        \label{fig:increamental_memory}
      }
      \vspace{-2em}
      \caption{(a) The operator's time proportion in multiple DLRM training jobs. (b) The memory demand of one job.}
      \vspace{-1.5em}
\end{figure}

Unlike traditional compute-intensive deep learning (DL) models used in computer vision (CV) and natural language processing (NLP), DLRM training incurs \textit{massive I/O operations} in addition to its compute-intensive operations (e.g., matrix multiplication for DNNs). These I/Os are largely due to frequent \textit{lookups} to embedding tables, consuming 30-48\% of the training time (see Fig.~\ref{fig:op_percentage}). Existing schedulers~\cite{Optimus, TetriSched, twine}, without considering such a unique blend of I/O and computation operations in the DLRM training, fall short in ensuring optimal resource utilization and training efficiency.

Another unique characteristic of DLRMs lies in that their embedding tables are notably {\it memory-intensive}. DLRMs can easily demand tens of terabytes of memory (\S\ref{subsec:DLRM_training_at_antgroup}). As user-targeted applications evolve, the size of embedding tables keeps increasing~\cite{memory_increase1, memory_increase3}.
For example, the memory usage of a typical DLRM job can surge to over 2.3TB within just 15 hours (see Fig.~\ref{fig:increamental_memory}). Consequently, there is a significant risk of hitting out-of-memory (OOM) for a DLRM job if the allocated resources cannot quickly adjust to its increased memory demand. We observed that, in the production environment of \company, thousands of jobs (5\% - 8\%) had been derailed due to OOM, leading to compromised user satisfaction and suboptimal cluster performance.

The high \textit{instability} of the cloud environment necessitates frequent scaling resources to adapt to the ever-changing cloud \cite{elasticity_ali,workload_consolidation_ali1,workload_consolidation_ali2}. To achieve this, traditional DL schedulers, like \cite{twine,Optimus}, require stopping a job 
%to disks 
and then restarting it with adjusted resources. This stop-and-restart process often takes up to tens of minutes (\S\ref{subsec:challenges_of_DLRM_training_at_antgroup}), highlighting the need for a more efficient approach. Additionally, model training with elastic training frameworks may result in inconsistent model accuracies due to stale gradients being submitted \cite{stale} or disruption of training data (\S\ref{subsec:challenges_of_DLRM_training_at_antgroup}).% \Hao{i rewrite this sentence. pls double check. -- looks much better: Hui} 

% the scheduling of DLRM training jobs even harder. 
% First, the job failure rate in the cloud is much higher than in dedicated clusters \cite{fault_in_the_cloud,failure_aware_system}. 
% For example, given a distributed job with 50 hundred components (e.g., hosted in pods on a Kubernetes platform), the failure rate could be as high as 53\% within a single day (\S\ref{subsec:challenges_of_DLRM_training_at_antgroup}).
% In addition, DLRM training jobs co-exist with other services (e.g., online services) with higher priorities. 
% These high-priority services can preempt resources from training jobs when they encounter workload spikes. This could result in the fail of DLRM training jobs or or the emergence of stragger (e.g., slow worker) due to insufficient resources. 

% More critically, as cloud resources are typically shared by multiple distinct services managed by different production teams, for isolation and security reasons, one does not know the resource usage of the service from another team, nor does he know the overall resource usage of the cloud. Therefore, DLRM's training can merely rely on the global information of cloud resources.
% These practical constraints also render state-of-the-art consolidation approaches \cite{twine, borg,borg2,workload_consolidation_ali1,workload_consolidation_ali2}, aiming to address performance issues from a global perspective, infeasible in training DLRMs in a dynamic, shared cloud environment.

To tackle these challenges, we introduce \system, a cloud-based deep learning training system designed for DLRMs.
%which attains exceptional throughput, high resource utilization, and robust fault tolerance.
%in the context of DLRM training. 
\system takes the runtime training information into account for accurately allocating and elastically scheduling resources for training jobs along with a bunch of novel mechanisms, including \textit{dynamic data sharding}, \textit{flash-checkpoint}, \textit{seamless migration}, and pre-adjustment-based, \textit{OOM prevention}. Together, \system attains exceptional throughput, high resource utilization, and robust fault tolerance.
%they ensure the robust execution of DLRM jobs in the cloud. 
In summary, we have made the following contributions:
%\Hao{I highlight the solution of \system here. please check.}

1) We build a \textit{resource-performance} model by considering I/O overhead and computation demands during DLRM training. With this model, we design a three-stage algorithm that can dynamically allocate resources during the whole cycle of DLRM training and significantly reduce the job completion time. 
%\Hao{I add "reduce the pending time", this is because the first stage algorithm in 3-stage algorithms aims to quick start the job. Please check the writing.} 
%We show that jobs with optimization could achieve comparable performance to well-tuned resource allocation.

2)  We invent a \textit{dynamic data sharding} mechanism to maintain the model quality  
when scaling or a job failure happens in the cloud. We further develop \textit{seamless migration} and \textit{flash-checkpoint} strategies to reduce the overhead of scaling jobs. We also develop an \textit{OOM-prediction} mechanism to prevent OOM.
%\Hao{we rewrite this contribution. please check it}  
%By employing these mechanisms, we  
%\Hao{This is not clear. Specifically: 1. what does "mitigate the perturbations" mean? 2. we need to connect our solution with mentioned challenge @Lan, pls rewrite the solution. }

3) We implement \system~ as a native auto-configuration service within Kubernetes and open-source all technical implementations. Thereby, end-users can train DLRM jobs in the production environment without concern for resource configuration and job failures. 

4) We thoroughly evaluate \system with thousands of jobs collected from months of various DLRM training workloads in a production environment equipped with more than 62K CPU and 3.24PB memory. The evaluation shows that \system improves the CPU utilization by 21.0\% to 27.6\% and memory utilization by 17.3\% to 31.6\%, and reduces job completion time by 30.9\% without compromising model accuracy.

% \begin{figure}[t] 
%     \centering
%     \hspace{-0.15cm}
%     \includegraphics[scale=0.125]{figures/increamental_memory.pdf}
%     \caption{The memory demand of a DLRM training job} 
%     %`\Lan{pls make this figure smaller and adjust the memory consumption of dense part to 1 GB.}}
%     \hspace{-0.15cm}
%     \label{fig:increamental_memory}
%     \vspace{-1.5em}
% \end{figure}

\vspace{-1em}
\section{Background and Motivation}
\label{sec:background_and_motivation}
%\Hao{I will rewrite this part.-10.10}

In this section, we briefly introduce DLRMs and the DLRM training platform at \company (\S \ref{subsec:DLRM_training_at_antgroup}). We then discuss the key issues when training DLRMs and the challenges to address them through investigating the unique characteristics of DLRMs and sharing the observations from our DLRM system deployed at \company (\S \ref{subsec:challenges_of_DLRM_training_at_antgroup}).

% \begin{figure}
%     \centering    \includegraphics{figures/fig2_recommendation_model_process.pdf}
%     \caption{Caption}
%     \label{fig:enter-label}
% \end{figure}

\begin{figure}[t] 
    \vspace{-1em}
    \centering
    \hspace{-0.15cm}
    \includegraphics[width=1\columnwidth]{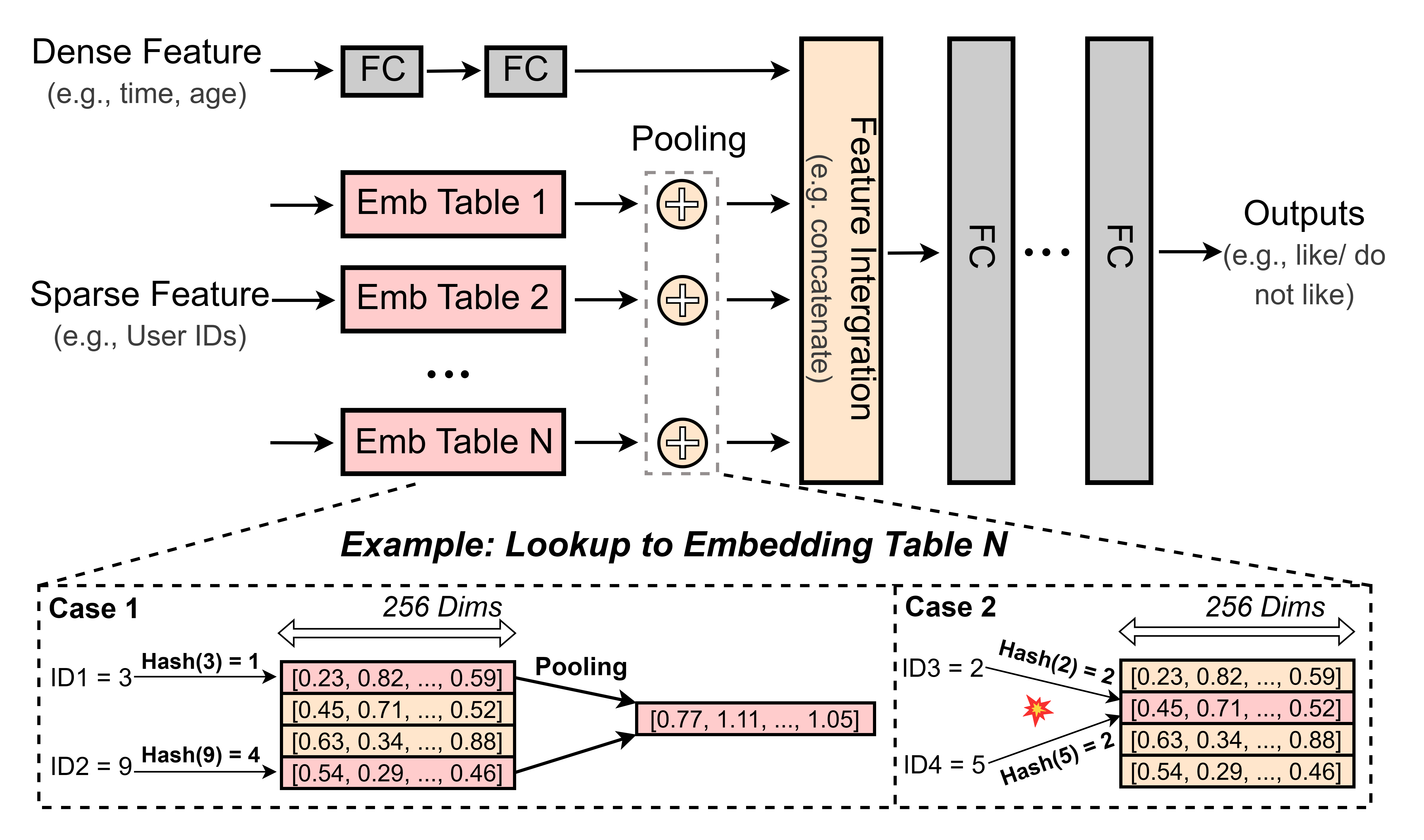}
    \vspace{-1.5em}
    \caption{ A typical DLRM architecture consists of neural networks, which make up the dense part, combined with memory-intensive embedding tables, forming the sparse part. The portion within the dashed box highlights examples of embedding table lookup in forward propagation of DLRM.}
    \label{fig:recommendation_model_overview}
    \vspace{-2em}
\end{figure} 

\vspace{-1em}
\subsection{DLRM Training at AntGroup}
\label{subsec:DLRM_training_at_antgroup}

At AntGroup, DLRMs are extensively applied to scenarios such as service/content search, marketing vouchers, Tab3 video recommendation, and advertising \cite{huan2023antm,ant_model_wd1}. Within our cloud-based cluster, DLRM training jobs account for more than 70\% of the total training jobs daily, consuming a significant amount of the cluster resources. 
Optimizing DLRM training holds substantial importance for the effective utilization of the cluster resources.

\noindent\textbf{Overview of DLRM.}
Fig.~\ref{fig:recommendation_model_overview} illustrates a typical model architecture of DLRMs: First, it integrates fully connected deep neural networks (DNNs) to capture continuous dense features like timestamps. Further, it uses embedding tables to transform various categorical sparse features, such as user and video IDs, into low-dimensional dense representations. These embedding tables represent the \textit{sparse part} of DLRMs, while the rest of the models (e.g.,  DNNs) represent the \textit{dense part}.
%In the forward pass of model computation, 
DLRMs take both sparse categorical and dense features as inputs for model training. 

As illustrated in Fig.~\ref{fig:recommendation_model_overview}, each categorical feature has its own embedding table. A data point, or an instance, for a categorical feature is mapped to a specific row in this table where its embedding vector is stored. For instance, consider a user with ID 3, representing an instance of the "User IDs" feature. The embedding for this user is located at the index ($hash(3)$ mod $M$) within the "User IDs" embedding table, which consists of $M$ rows.
%each feature instance (e.g., a specific user ID of feature "User IDs") is used to look up a unique row (e.g., embedding vector) in the embedding table. 
%For example, given a user with an ID $3$ (i.e., an instance of "User IDs" feature), its embedding is stored at index ($hash(3)$ mod $M$) of the embedding table (of $M$ rows) for the "User IDs" feature.
%of an embedding table with $M$ rows (see Fig.~\ref{fig:recommendation_model_overview}).
Embedding vectors that are being accessed undergo an integration process into a singular dense vector through element-wise \textit{pooling} operations. These operations typically use either the sum or maximum value of the vectors. This pooled embedding vector is further concatenated with the intermediate output derived from the dense features. This combined output forms the input for the subsequent DNN layers. In practice, a DLRM might comprise thousands of embedding tables, with some tables having millions of rows~\cite{AdaEmbed}.
%The accessed embedding vectors are further integrated into a single dense vector using element-wise \textit{pooling} operations, which typically take \textit{sum} or \textit{maximum} for the vectors. Then the pooled embedding vector concatenates together with the intermediate output of the dense features, forming the input for the deeper DNN layers. In practice, a DLRM might have thousands of embedding tables, some of which contain millions of rows \cite{AdaEmbed}.

%The accessed embedding vectors are then integrated into a single dense vector using element-wise \textit{pooling} operations, which typically take \textit{sum} or \textit{maximum} for the vectors. Then the pooled embedding vector concatenates together with the intermediate output of the dense features, forming input for deeper DNN layers. In practice, a DLRM might house thousands of embedding tables and some of them contain millions of rows. \cite{AdaEmbed}

%In the backward pass, the weights of the accessed embeddings are updated using the gradient.

%\noindent\textbf{Training architecture of DRLMs.}
% Typical parallel mechanisms in deep learning include \textit{data parallelism} and \textit{model parallelism} \cite{large_minibatch_sgd}. Data parallelism divides training data into multiple datasets for each worker, while model parallelism partitions the models into multiple parts that can be trained in parallel. 

\begin{table}[ht]
\label{tab: Cost comparison between GPU and CPU}
\centering
\vspace{-1.3em}
\caption{DLRM training cost comparison between the hybrid(GPU-CPU) and CPU-only approach on AWS. Sample/price represents the number of data points that can be trained per unit cost, measured in millions per USD.}
\vspace{-0.15in}
\label{table:cpu-gpu compare}
\resizebox{0.45\textwidth}{!}{%
\begin{tabular}{|l|c|c|c|l|l|l|}
\hline
\multirow{2}{*}{\textbf{Model}} &
  \multirow{2}{*}{\textbf{Device}} &
  \multirow{2}{*}{\textbf{Time}} &
  \multirow{2}{*}{\textbf{\begin{tabular}[c]{@{}c@{}}Unit\\ Price\end{tabular}}} &
  \multicolumn{1}{c|}{\multirow{2}{*}{\textbf{\begin{tabular}[c]{@{}c@{}}Samples/\\ Price\end{tabular}}}} &
  \multicolumn{1}{c|}{\multirow{2}{*}{\textbf{\begin{tabular}[c]{@{}c@{}}CPU\\ Util\end{tabular}}}} &
  \multicolumn{1}{c|}{\multirow{2}{*}{\textbf{\begin{tabular}[c]{@{}c@{}}GPU\\ Util\end{tabular}}}} \\
 &
   &
   &
   &
  \multicolumn{1}{c|}{} &
  \multicolumn{1}{c|}{} &
  \multicolumn{1}{c|}{} \\ \hline
\multirow{2}{*}{\begin{tabular}[c]{@{}l@{}}Wide\&\\ Deep\end{tabular}} &
  CPU &
  1.41h &
  0.53usd/h &
  {\ul 3.4m/usd} &
  $\approx$ 33\% &
  \diagbox[width=3em]{}{}
   \\ \cline{2-7} 
 &
  Hybrid &
  0.98h &
  3.59usd/h &
  {\ul 1.9m/usd} &
  $\approx$ 26\% &
  $\approx$ 3\% \\ \hline
\multirow{2}{*}{DeepFM} &
  CPU &
  1.53h &
  0.53usd/h &
  {\ul 3.1m/usd} &
  $\approx$ 34\% &
  \diagbox[width=3em]{}{}
   \\ \cline{2-7} 
 &
  Hybrid &
  0.95h &
  3.59usd/h &
  {\ul 2.1m/usd} &
  $\approx$ 28\% &
  $\approx$ 4\% \\ \hline
\end{tabular}
}

\vspace{-1em}
\end{table}

At AntGroup, we use the parameter server (PS) architecture ~\cite{ps-framework} for DLRM training as it serves as a {\it de facto} framework for this purpose \cite{mudigere2022software,zhao2019aibox,gupta2021training,acun2021understanding,hierarchical_training}.
% \Hao{this sentence is not good. I will rewrite this.}
% It consists of three nodes: trainers, sparse parameter servers (sparse PSes), and dense parameter servers (dense PSes) \cite{hierarchical_training}. The embedding tables, which form the sparse part of the model, are distributed across the {\it sparse PSes}, shared by all trainers. These sparse PSes can be scaled up by incorporating more machines, especially when the number of embedding tables increases. In contrast, the dense part of the model is duplicated across each trainer. It means that every trainer maintains a local copy of the dense parameters. During training, each trainer accesses and modifies its local dense parameters. Synchronization of these dense parameters across all trainers is conducted in the background by the dense PSes.

\noindent\textbf{Hardware Selection: CPU-only.} In practice, there are two approaches in training recommendation models: 1) a {CPU-only} approach by exclusively using CPUs; 2) a {CPU-GPU} hybrid approach by employing CPUs for handling the embedding data while GPUs for executing data-parallel neural networks~\cite{memory_increase2}. While the CPU-GPU approach is the preferred choice for companies like Meta \cite{acun2021understanding, AdaEmbed}, we lean toward the CPU-only approach due to the following reasons: First, CPU resources are more readily accessible and cost-effective compared to GPUs in a production environment -- our clusters have an abundance of CPU resources, e.g., over 200k cores, whereas housing fewer than 1,000 high-performance GPU cards. 
% \Hao{Considering that the majority of GPUs at \company are now dedicated to serving Gen AI, the available GPUs for DLRM training have become more limited.} 
Further, given the scarcity of GPU resources, it is desired to maximize their utilization. 
%However, the CPU-GPU hybrid training for recommendation models involves intensive I/O operations:
However, DLRM training involves intensive I/O operations, including 
1) transferring embedding data between CPUs and GPUs (up to 22\%~\cite{memory_increase2} of the total training time), and 2) conducting a myriad of lookup operations to the embedding tables (over 30\%  of the total training time). Such massive I/O operations render GPUs underutilized. As shown in Table \ref{table:cpu-gpu compare}, when training the two most common DLRM models at \company \cite{ant_model_wd1,ant_model_wd2,ant_model_dcn1,ant_model_wd1}: 1) the average GPU utilization under the CPU-GPU hybrid approach is lower than 3\%, while 2) the CPU-only approach can train more data at a unit price. Therefore,  \uline{we use CPU to train DLRMs and base the remaining discussion on the CPU-only approach}. 

% Our observation indicates that the CPU-GPU training not only results in inefficient use of valuable GPU resources but also impacts the overall speed and efficiency of model training. 

\noindent\textbf{Cloud Environment: Workload Consolidation.} Table \ref{tab:statistic-of-jobs} shows that at AntGroup, different types of jobs (i.e., training, serving, and stream processing jobs) are running in the same cluster and sharing the resources \cite{sang2023cougar}. For isolation and security reasons, the DLRM system is unaware of the resource usage by other
 services, as well as the overall resource consumption of the cloud. That said, the DLRM system has no direct control over the cluster resources and has to request resources from the cluster resource scheduler when scheduling or scaling out a
 DLRM training job. 

% \begin{figure}[t] 
%     \centering
%     \hspace{-0.15cm}
%     \includegraphics[scale=0.125]{figures/op_percentage.pdf}
%     \caption{The proportion of calculation time for DLRMs.} 
%     \hspace{-0.15cm}
%     \label{fig:op-percentage}
%     \vspace{-1.5em}
% \end{figure} 

% \begin{table}[]
% \label{tab: Cost comparison between GPU and CPU}
% \begin{tabular}{|l|c|c|c|c|c|}
% \hline
%                              &          &         &       & \multicolumn{1}{c|}{}                    \\
% \multirow{-2}{*}{\textbf{Model}} &
%   \multirow{-2}{*}{\textbf{Device}} &
%   \multirow{-2}{*}{\textbf{\begin{tabular}[c]{@{}c@{}}Samples/ \\ Price\end{tabular}}} &
%   \multirow{-2}{*}{\textbf{\begin{tabular}[c]{@{}c@{}}CPU \\ Util\end{tabular}}} &
%   \multirow{-2}{*}{\textbf{\begin{tabular}[c]{@{}c@{}}GPU \\ Util\end{tabular}}} \\ \hline
%                              & CPU      & 99m/usd & $\approx$ 41\% & \diagbox[width=3em]{}{} \\ \cline{2-5} 
% \multirow{-2}{*}{Wide\&Deep} & CPU+V100 & 30m/usd & $\approx$ 33\% & $\approx$ 3\%             \\ \hline
%                              & CPU      & 99m/usd & $\approx$ 34\% & \diagbox[width=3em]{}{} \\ \cline{2-5} 
% \multirow{-2}{*}{DCN}        & CPU+V100 & 30m/usd & $\approx$ 30\% & $\approx$ 4\%             \\ \hline
% \end{tabular}
% \captionsetup{labelsep=colon} % 设置特定表格的标题分隔符为冒号
% \caption{DLRM training cost comparison between GPU and CPU on AWS.\Lan{Sample/price meaning?}}
% \label{table:cpu-gpu compare}
% \end{table}
% % \Lan{m means million, sample cite criteo}

% Please add the following required packages to your document preamble:
% \usepackage{multirow}

\begin{table}[h]
\vspace{-1em}
\centering
\caption{Statistic of Jobs at \company}
\vspace{-0.15in}
\label{tab:statistic-of-jobs}
\resizebox{0.45\textwidth}{!}{%
\begin{tabular}{|l|c|c|c|c|}
\hline
\textbf{Job Type}  & \textbf{Count} & \textbf{vCPU} & \textbf{CPU Util} & \textbf{MEM}  \\ \hline
Training          & 62K            & 600K       &20\%         & 0.9PB              \\ \hline
Stream Processing & 43K            & 450K       &15\%         & 0.63PB       \\ \hline
Inference Service & 3K             & 300K       &10\%         & 0.41PB             \\ \hline
Search Service    & 0.9K           & 200K       &15\%         & 1.2PB              \\ \hline
Other             & 2K             & 50K        &10\%         & 0.1PB              \\ \hline
\end{tabular}
}
\vspace{-1em}
\end{table}

\vspace{-1em}
\subsection{Challenges of DLRM Training at AntGroup  }
\label{subsec:challenges_of_DLRM_training_at_antgroup}

By collecting significant training task data from the largest machine learning platform at AntGroup, we identified two primary issues with DLRM training: {\it low resource utilization} and {\it high cloud instability}. In this section, we provide a detailed analysis of these two problems and highlight the challenges in addressing them.
%Moreover, we emphasize the challenges encountered in resolving these issues.

\begin{figure}[t]
\vspace{-1em}
    \hspace{-2em}
    \label{fig:job_stats}
    \subfigure{
        \includegraphics[scale=0.125]{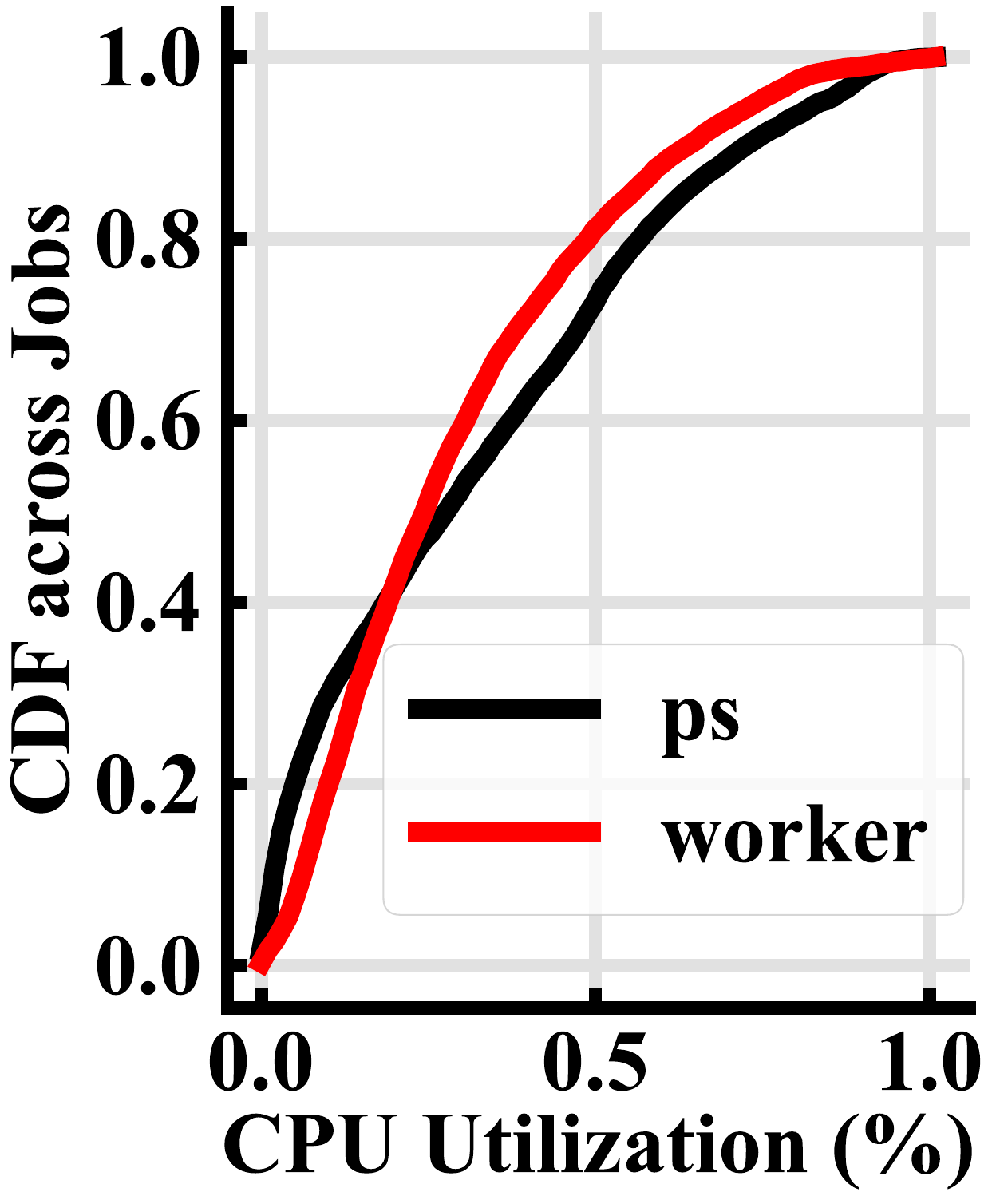}
        \label{fig:job_cpu_stats}
    }
    % \hspace{-1em}
    \subfigure{
        \includegraphics[scale=0.125]{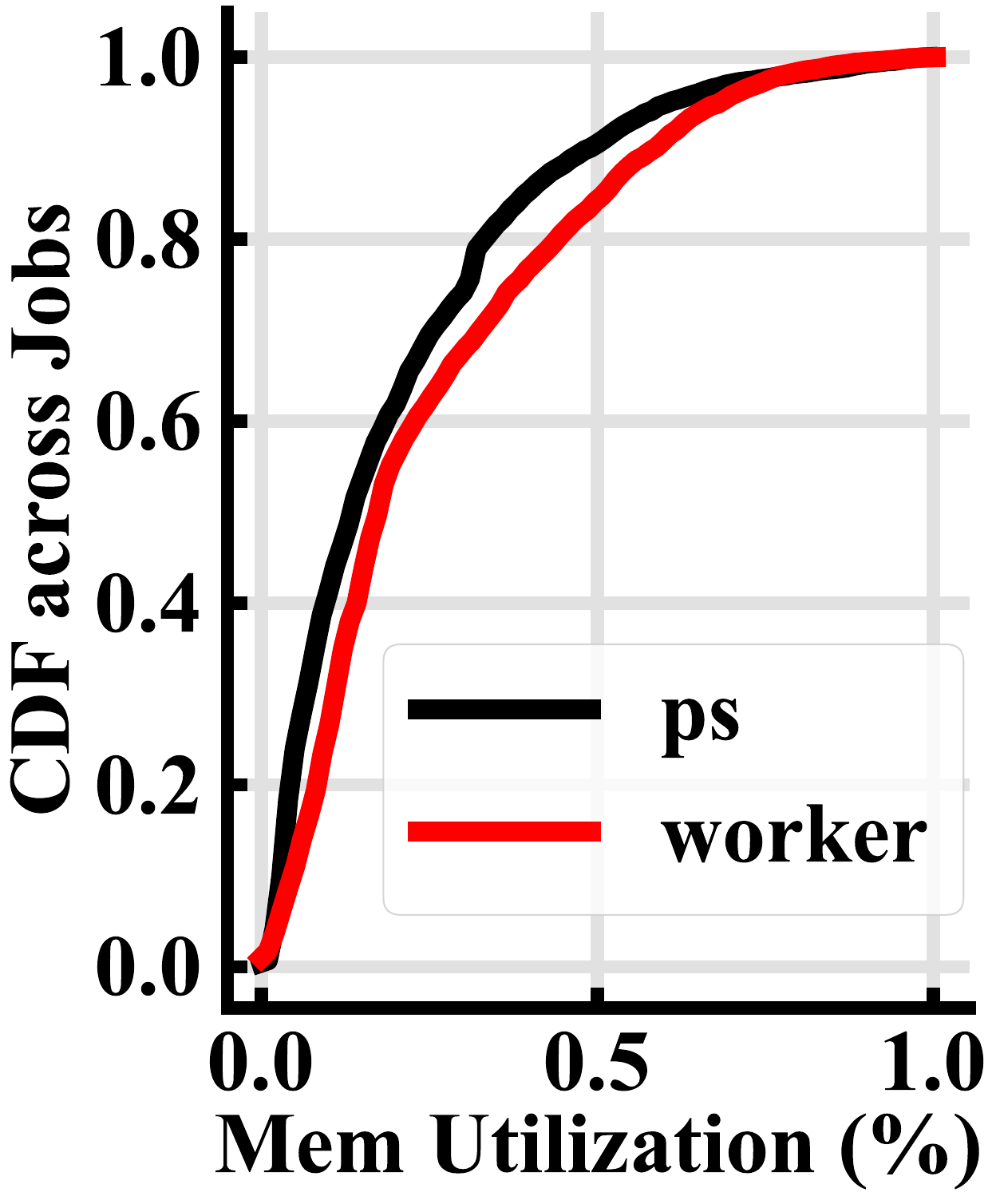}
        \label{fig:job_memory_stats}
    }
    % \hspace{-1em}
    \subfigure{
        \includegraphics[scale=0.125]{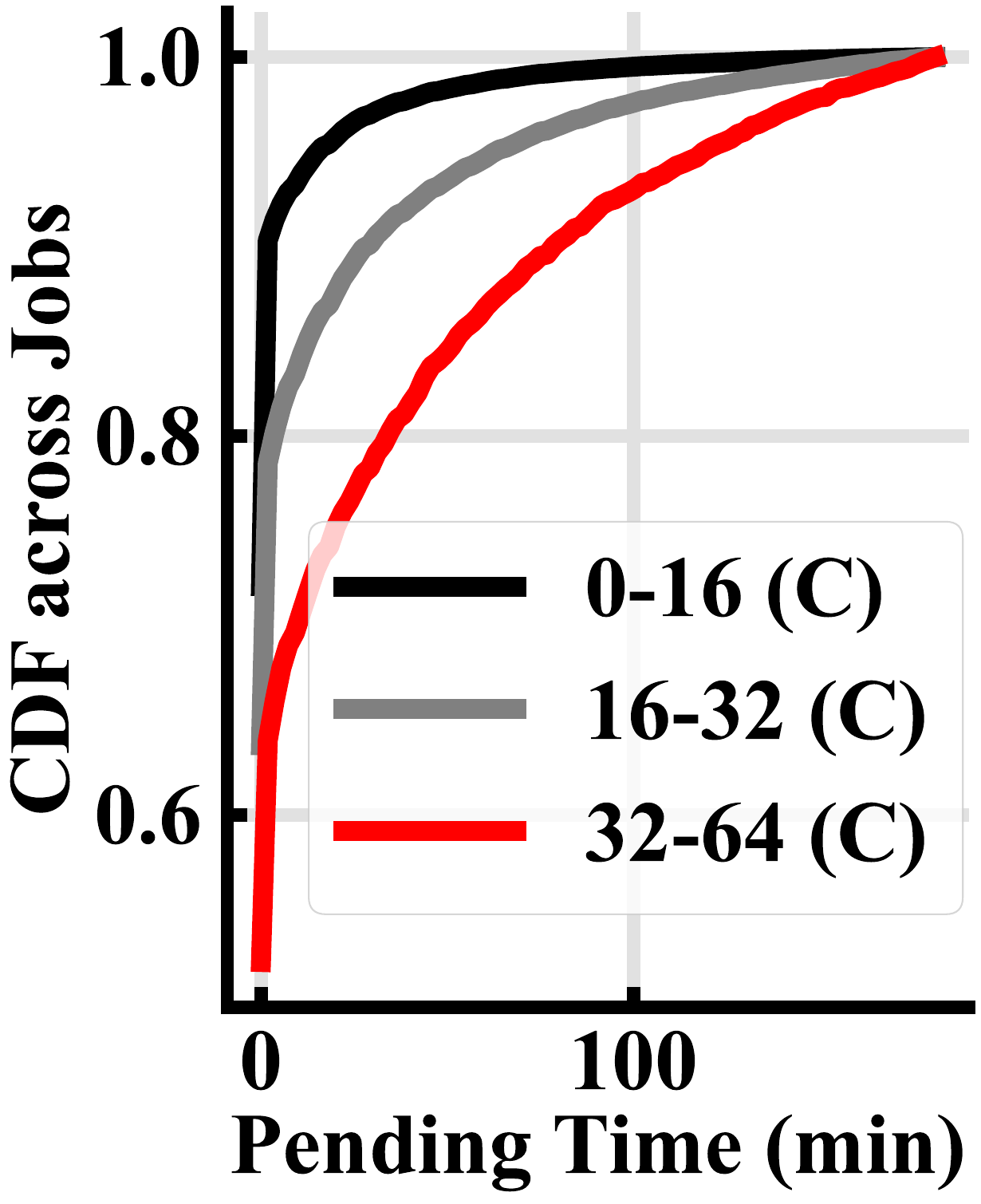}
        \label{fig:pod_pending_time}
    }
    \hspace{-2em}
    \vspace{-0.2in}
    \caption{DLRM jobs' resource utilization and pending time derived from cluster traces in \company.}
    \label{fig:job_resource_configuration_before}
    \vspace{-2em}
\end{figure}

 \noindent\textbf{Low Resource Utilization.}  As depicted in Fig.~\ref{fig:job_resource_configuration_before}, over 80\% of the jobs in our cluster had CPU and memory utilization rates below 50\% back in 2021, resulting in a significant waste of cluster resources. The core factor causing this is \textit{suboptimal configurations by users}. Specifically,  in a typical cloud environment, cloud users need to specify a fixed amount of resources before deploying their cloud-based services \cite{kubeflow2023,borg,twine}.
%\Lan{pls add citation}.
%other deep learning systems~\cite{tensorflow2015-whitepaper},
Similarly, our previous training system running in a cloud-based cluster also needed such inputs, i.e., resource configurations, from system users (e.g., ML engineers or data scientists). Such resource configurations were used to guide the training system for resource allocation during DLRM training.  Users typically resorted to a time-consuming \textit{trial-and-error} approach to determine these configurations -- by manually (re-)running their jobs multiple times with varying resource configurations in search of the "optimal" one. Oftentimes, these user-provided configurations tended to ask for "more-than-needed" resources to avoid job failures during training, resulting in inefficient use of resources. 

To overcome this, instead of relying on user-provided suboptimal resource configurations, we need an approach allowing the DLRM system to automatically allocate and dynamically adjust resources for training jobs for high resource utilization. Note that for distributed training, resource adjustment includes changing the number of nodes (horizontal scaling) and the resources of each node (vertical scaling)~\cite{borg}. This is nontrivial due to two main challenges:

\vspace{-0.5em}

\begin{itemize}[noitemsep,leftmargin=10pt]

    \item \textbf{Timely Meeting Memory Demands.} The memory requirement for storing embedding tables can surge up to 10s of terabytes~\cite{AdaEmbed} in a short period. As shown in Fig.~\ref{fig:increamental_memory}, the memory usage of embedding tables in a typical DLRM model can spike to more than 2.3TB within 15 hours. This renders DLRMs vulnerable to out-of-memory (OOM) issues if memory allocation cannot timely meet the model's demands -- we observed 5\%-8\% of jobs suffering from OOM in our production environment, greatly affecting the overall cluster resource efficiency.
    
    \item \textbf{Precisely Allocating CPU Resources.} In DLRM training, extensive I/O operations impact both model training efficiency and job CPU utilization. As shown in Fig.~\ref{fig:op_percentage}, the lookup operations can account for  30\%-48\% of the training duration in a single iteration. Conventional deep learning resource schedulers \cite{Optimus, TetriSched, twine} fall short of handling this unique training process, often overlooking the lookup latency and allocating inappropriate CPU resources for training jobs. 
% In this study, our emphasis lies on {optimizing the scheduling of training jobs' CPU resources to enhance overall CPU utilization}.

\end{itemize}

\vspace{-0.5em}

\noindent\textbf{High Cloud Instability.} Unlike a dedicated cluster (for a single-purpose service), the cloud environment has a much higher job failure rate \cite{fault_in_the_cloud, failure_aware_system}. Statistically, we observed that the daily failure rate for a simple job (e.g., hosted in a single Kubernetes pod) in our cloud-based cluster is 1.5\% due to network errors, node malfunction, etc. The failure rate increases exponentially for a more complex distributed job with hundreds of components. For example, the daily failure rate for a job with 50 pods
increases dramatically to $1~-~(1-0.015)^{50} = 53.03\%$. 
Moreover, in our cloud-based cluster, different services co-exist, sharing the same cloud resources (\S \ref{subsec:DLRM_training_at_antgroup}). Compared to other higher-priority services, e.g., online services, DLRM training is typically labeled with a lower priority. When higher-priority services encounter workload spikes, the cluster scheduler preempts resources allocated to the DLRM system, resulting in the failure of DLRM training jobs or the emergence of stragglers (e.g., slow workers) due to insufficient resources.

To address this, our system needs the capability to 1) frequently scale up/down training jobs to adapt to the changing cloud environment and 2) detect failed nodes and recover them swiftly. However, this is also not trivial due to the following two main challenges:

\begin{itemize}[leftmargin=10pt]
    \item \textbf{Ensuring Consistent Model Quality.} 
    Elastic training frameworks can enhance training throughput by dynamically scaling training jobs up or down. However, elasticity operations (e.g., increasing/decreasing the number of worker nodes and/or adding/shrinking computational resources to a worker) can also lead to inconsistent job configurations (e.g., batch size and the number of parallel workers) and/or changed data sequences. For example, some slow workers may submit too many stale gradients to PSes, causing instability in parameter updates; some workers might miss specific data batches due to failures, or the training data sequence could be disrupted during scaling operations (\S \ref{subsec: dynamic_data_sharding}). These inconsistent configurations and disruptions could further compromise the consistency of model training quality~\cite{stale,easyscale}, especially in asynchronous training \cite{chen2017revisiting}.
    
    %Although elastic training frameworks can improve training throughput by scaling up/down training jobs, the inconsistency in job configuration (e.g., batch size and the number of parallel workers) may lead to inconsistent model quality~\cite{stale,easyscale}, especially in asynchronous training \cite{chen2017revisiting}. For example, some slow workers may submit too many stale gradients to PSes, causing instability in parameter updates. In addition, data inconsistency may also lead to an inability to guarantee model quality. For example, workers might miss specific data batches due to worker failures, or the training data sequence could be disrupted during scaling operations (\S \ref{subsec: dynamic_data_sharding}). These disruptions could compromise the consistency of model training quality. \Hao{I rewrite this paragraph. pls double check.}
    
    \item \textbf{Providing Fast Elasticity.} 
    Swift scaling operations are essential for accelerating jobs (e.g., by allocating more resources) and managing instability (e.g., by addressing slow workers in a job group). Conventional DL schedulers~\cite{Optimus,lyra} involve a {\it stop-and-restart} operation to scale up/down a job -- by saving the job's checkpoint (to hard disks) and restarting the job with adjusted resources/configurations, e.g., reallocating training data (to scale workers) or re-partitioning the model (to scale PSes). The {\it stop-and-restart} operation is very {\it costly}: First, checkpointing a job to remote disk storage (RDS) typically takes 5-10 minutes~\cite{wang_gemini_2023}. Further, the scheduler takes another 5-10 minutes to complete the necessary preparation before restarting, including submitting a new job \texttt{YAML}, requesting resources for the new pods, pulling images from the registry, and re-establishing the code environment. Under conditions of resource scarcity (e.g., daytime \cite{workload_consolidation_ali2}), the duration can extend beyond 30 minutes. Last, loading the checkpoint from RDS and restarting the training takes another 5-10 minutes. Altogether, the whole process could consume tens of minutes, introducing high overhead for DLRM training (\S \ref{subsec:seamless_migration}).

    %In cloud environments, frequent scaling is essential for both job acceleration (e.g., re-allocating resources to speed up jobs) and instability handling (e.g., handling slow workers in job groups). In conventional DL schedulers like \cite{Optimus,lyra}, scaling up/down a job needs to save its checkpoint to hard disks and restart the job, as they need to reallocate training data (to scale workers)/ re-partitioning the model (to scale PSes). The whole process involves three parts of overheads. First, the job needs to save checkpoint to remote disk storage (RDS), consuming around 5-10 minutes \cite{wang_gemini_2023}. Further, the scheduler needs to initilize the necessary preparation before restarting, including submitting a new job \texttt{YAML}, requesting resources for the new pods, pulling images from the registry and re-estabilishing the code environment.  This process typically takes around 5-10 minute, but it can extend to over 30 minutes during periods of resource scarcity (e.g., daytime \cite{workload_consolidation_ali2}). Last, the job needs to load checkpoint from RDS and restart training (5-10 minutes). Altogether, the whole process could consume tens of minutes, introducing high overhead for DLRM training (\S \ref{subsec: seamless_migration_and_OOM_handling}).  \Hao{I rewrite this paragraph. pls double check. At the same time, i wanna address the importance  of `fast checkpoint` mechanism. Should we write this here or in design objective(next section)}

\end{itemize}

%are sharing the resources of the cluster to win the maximum resource utilization. Compared to online services, DLRM training is marked as a lower priority. When user access to the online service spikes, the cluster scheduler preempts resources allocated to DLRM scheduler to maintain the Quality of Service (QoS) for high-priority jobs. This usually ends up with the failure of one or more DLRM training jobs.

%\noindent \textbf{Computation Analysis.}
\vspace{0.05in}

%In this work, we focus on \uline{training job CPU resource optimization ++++++++++++++++++++++++++++++++++++++++++++++++scheduling to improve the CPU utils}. 

%Due to the inherent complexity of DLRM as introduced, previously, ML engineers or Data scientists need to configure the DLRM jobs resource manually. The simple but most commonly employed strategy is \textit{trial-and-error }. In other words, users manually initiate the same job repeatedly with different resources to search for a good resource configuration. This is very time-consuming since a new job needs to launch Pods, prepare the runtime environment, and initialize the computation graph. Users also have to wait for a while before they can observe useful information, such as the job's throughput and resource utilization. As shown in Fig.  \ref{fig:job_stats}, more than 80\% job’s CPU and memory utilization is below 50\%  at our cluster in 2021. Worse more, even if the cluster already provides more than 20K CPU cores and 500TB RAM resources, the pending time is considerably long for new jobs requiring a large number of resources (e.g. 32-64 CPU cores) to train large-size models.   

%\noindent\textbf{Handcrafted resource allocation is suboptimal.} 

%At the same time, thousands of jobs with this method resources may still deviate considerably from the optimal resource configuration, leaving room for improving resource utilization in clusters with a tailor-made resource scheduler. 

%\Hao{TODO-10-11: replot the figure2 and 3, ensemble them togeter. }
%\label{subsec: challenge for DLRM}
\begin{figure*}[ht]
\vspace{-2em}
\label{fig:dlrover-overview}
\centering
\includegraphics[width=2.2\columnwidth]{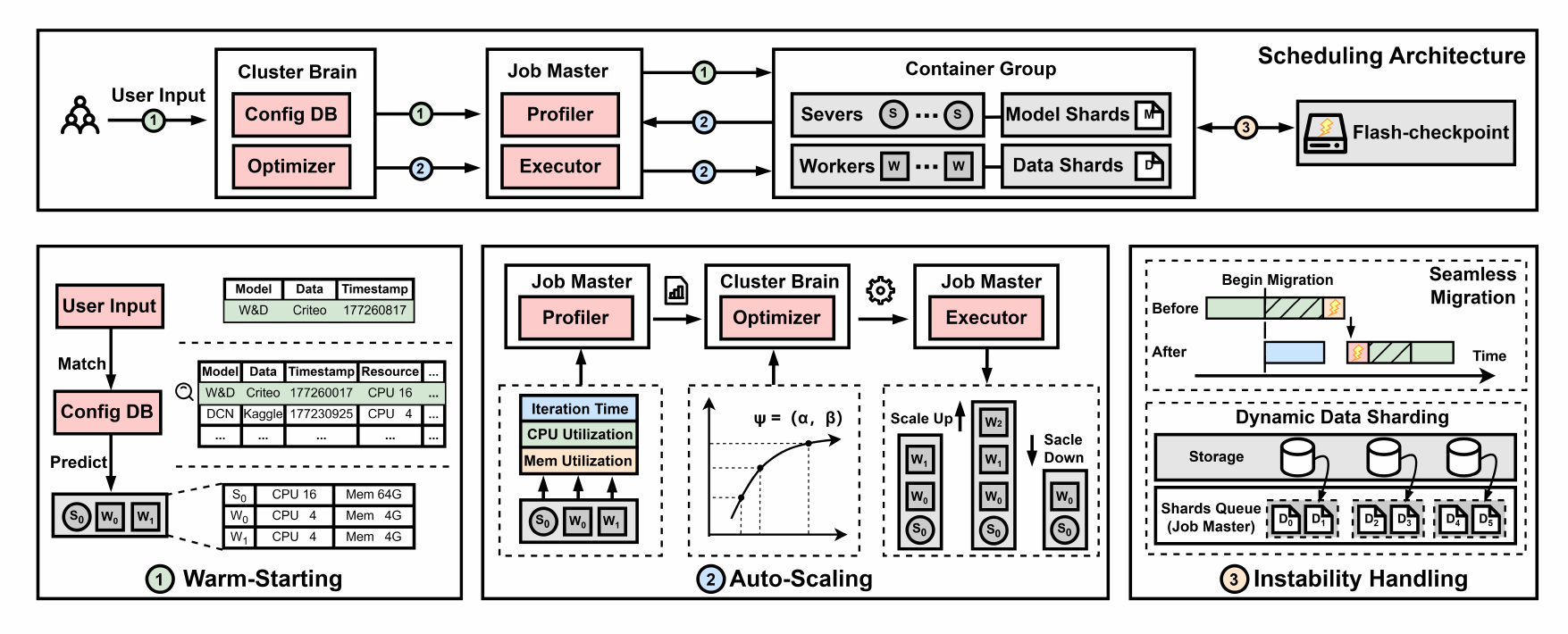}
\vspace{-3em}
\caption{Overview of DLRover-RM and Model Training Workflow}
\vspace{-1em}
\label{fig:system_dataflow}
\end{figure*}

\vspace{0.05in}

\vspace{-1.5em}
\section{Overview of DLRover-RM}
\label{sec:overview_of_dlrover}
In this section, we present the architecture overview of \system and highlight its key design objectives.

\vspace{0.05in}
\noindent{\bf Design Objectives.}
\system focuses on efficiently training a multitude of DLRMs simultaneously in a dynamic, shared cloud environment. It dynamically schedules computational and memory resources for DLRM training jobs to optimize training throughput and resource utilization while mitigating the job failure rate. The design objectives of \system are to answer the two key questions:
%\system is crafted to proficiently train thousands of DLRMs in a cloud setting with constrained resource allocation privileges. Its primary function is to judiciously allocate computational and memory resources, maximizing training throughput and optimizing resource utilization. At the same time, it is robustly designed to diminish the likelihood of DLRM job failures, ensuring consistent and uninterrupted training sessions in these restricted environments. Thus, the design of \system addresses two key objectives: 
%\Hao{If we can highlight the DLRover is designed for real-world production-level cluster here?}
\begin{itemize}[leftmargin=10pt]
    \item \textbf{Automated Resource-Performance Optimization:} 
    How can \system accurately allocate resources for DLRM training jobs -- without user-handcrafted configurations -- to maximize training throughput and minimize resource cost (\S\ref{sec:exploring_optimal_configurations})
    
    \item \textbf{High Stability Assurance:} How does \system overcome the dynamic nature of the cloud environment to achieve robust execution of DLRM training jobs with low job failure rate and high fault tolerance (\S\ref{sec:handling_instability})
\end{itemize}

\noindent{\bf Architecture Overview.} \system is based on the parameter
server architecture (see \S\ref{subsec:DLRM_training_at_antgroup}) in our production cloud environment. As illustrated in Fig.~\ref{fig:system_dataflow}, \system consists of two main components: 1) a cluster-level central coordinator, called {\it cluster brain}; and 2) a group of job-level distributed training agents, called {\it job master}:
%We designed \system based on the parameter server architecture, as described in the \textit{Parameter Server} framework \cite{ps-framework}, within a production-level cloud environment. The \textit{AllReduce} framework, often used in distributed training, did not align with our production requirements due to two critical constraints: (1) The performance of synchronous frameworks, such as \textit{AllReduce}, is notably affected in the cloud environment, primarily due to the higher failure rate \cite{ps-for-training}. Cloud environments are known for their dynamic nature, and the increased likelihood of node failures can impact the reliability and speed of synchronous training approaches. (2) In our specific cloud setup, we have an abundance of CPU resources but limited GPU resources, with, for example, 1.5 million CPUs compared to 4.5k GPUs. This resource imbalance poses challenges when using GPU-intensive frameworks like \textit{AllReduce}, as it may not efficiently utilize the available CPU resources. To address these constraints, we opted for the \textit{Parameter Server} architecture as a more suitable solution for our cloud-based deep learning training needs, providing greater flexibility and resilience in handling the cloud environment's characteristics, ensuring efficient and reliable training.
%\Tang{update the citation here}
 
%\system comprises of a cluster-level central coordinator \textbf{cluster brain} and a group of job-level distributed training agents \textbf{job master} as Fig.~\ref{fig:system_dataflow}:
 
\begin{itemize}[leftmargin=10pt]
    \item The \textbf{cluster brain} comprises two subcomponents: the \textit{optimizer \text{and} config database (config DB)}. The \textit{optimizer} receives the runtime profiles (e.g., CPU and memory utilization) of training jobs from each \textit{profiler} periodically. With such information, the {\it optimizer} creates resource plans and sends them to the corresponding \textit{executors}. Meanwhile, the \textit{config DB} stores the information as the historical job traces. 
    
    \item Each \textbf{job master} also comprises two subcomponents: the \textit{profiler} and \textit{executor}. The \textit{profiler} monitors and collects runtime information for each job (i.e., from its workers and PSes) in a fixed interval and reports it to the {\it optimizer} of the \textit{cluster brain}. The \textit{executor} feeds data shards (e.g., a slice of training data) to the job's workers (e.g., hosted in pods) for training.

\end{itemize}

\noindent\textbf{Life Cycle of Training}. 
As detailed in Fig.~\ref{fig:system_dataflow},  upon submission of a job by the user, the \textit{cluster brain} quickly learns the job's characteristics -- by leveraging relevant historical data from the \textit{config DB} -- and then generates an initialization (\textit{warm-starting}) resource plan with the relative configuration (e.g., the number of CPU for each worker/PS) and similarity information (e.g., time series information)(\blacknumber{1}). Note that, at this moment, we choose a reasonable configuration near the optimal configuration (hence, with fewer scaling operations and shorter scaling times for auto-scaling) instead of pursuing an optimal configuration. 
Subsequently, the \textit{cluster brain} sends the \textit{warm-starting} resource plan to the respective \textit{job master} for job initialization. 

During job running, the \textit{profiler} profiles the job's runtime statistics and reports them back to the \textit{optimizer} periodically. With such updated runtime information,  the \textit{optimizer} can generate a refined resource plan, upon which the \textit{executor} dynamically adjusts the number of workers and/or PSes, and their resource configurations accordingly, i.e., the execution plan (\blacknumber{2}). 
%\Hao{check the writing in the third step of workflow @Dr.Lv -- looks good now.}
% Meanwhile, the \textit{executor} needs to redistribute the data shards to ensure the model quality (\S \ref{subsec: dynamic_data_sharding}).
%during training through a novel \textit{dynamic data sharding} mechanism 
% ~\footnote{
% It ensures that the adjustment of resources does not hurt the performance of the model (i.e., accuracy and convergence). For instance, suppose that there are two workers, each with a mini-batch size of 32. In this case, the global batch size is 64. Now, if the number of workers is increased from 2 to 4, the mini-batch size for each worker is adjusted to 16 to ensure the same global batch size of 64. 
% }. 
%and the data manager redistributes the data shard with the corresponding size. 
%Note that, the job will continue to train models without pending when resource requirements for scale-up can not be filled immediately. 
%The scale-up will happen later when all required resources are ready.  
%of previous steps of the workflow. 

\system further provides a set of reliable instability handling mechanisms to ensure the robust execution of training jobs (\blacknumber{3}). For failed/slow workers, \system implements a \textit{dynamic data sharding} mechanism to redistribute missed data and rebalance workloads between workers (\S\ref{subsec: dynamic_data_sharding}). For failed/slow PSes, \system devises a \textit{seamless migration} with in-memory checkpoint, named \textit{flash-checkpoint}, to minimize the overhead in failure recovery and job migration (\S\ref{subsec:seamless_migration_and_OOM_handling}). We will look closer at this three-stage job life cycle of \system (\blacknumber{1}-\blacknumber{3}) in our three-stage algorithm (\S\ref{subsec:3-stage_auto-scaling_optimization three-stage-algorithm}).

\vspace{-1em}
\section{Exploring Optimal Configurations}
\label{sec:exploring_optimal_configurations}

In this section, we present how \system accurately allocates resources for DLRM training. It first builds a resource-performance model 
%(i.e., \textit{cpu-throughput} model) 
for DLRM training (\S\ref{subsec:resource-performance_modeling}). Then it formulates the optimizing objective (\S\ref{subsec:optimization_formulation}). Finally, it proposes a novel three-stage algorithm to guide resource allocation (\S\ref{subsec:3-stage_auto-scaling_optimization three-stage-algorithm}).
%We first build the performance model (i.e., \textit{cpu-throughput} model) for DLRM training (\S\ref{subsec:resource-performance_modeling}), followed by formulating our optimizing objective (\S\ref{subsec:optimization_formulation}). Lastly, we propose a novel three-stage algorithm to guide resource allocation (\S\ref{subsec:3-stage_auto-scaling_optimization three-stage-algorithm}).
%in a stepwise manner,

\begin{table}[h]
    \vspace{-0.5em}
    \renewcommand\arraystretch{1.2} % Reduce from 1.4 to 1.2
    \centering
    \caption{Set of Common Notations.}
    \vspace{-0.15in}
    \label{table: Frequently used notations}
    \small % Reduce font size for the table
    \begin{tabular}{>{\bfseries}l l} \hline
    \textbf{Notation} & \textbf{Definition} \\ \hline
    $m$ & Batch size  \\
    $w/p$ & \# of \textit{workers}/\textit{parameter servers}  \\
    $\lambda_{w}/\lambda_{p}$ & CPU allocation of a \textit{worker}/\textit{parameter server} \\
    % $\rho_{w}/\rho_{p}$ & Memory consumption of a \textit{worker}/\textit{parameter server} \\
    % $\lambda_{\text{limit}}/\rho_{\text{limit}}$ & Total CPU/memory capacity \\
    $\Psi_{thp}$  & Training job throughput \\
    $a_{r}$  & Resource allocation for resource $r$ \\
    $A$  & The set of resource allocation for all resources \\
    \hline
    \end{tabular}
    \vspace{-1.5em}
\end{table}

\vspace{-1em}
\subsection{Resource-Performance Modeling}
\label{subsec:resource-performance_modeling}
% We model the computation performance of DLRM training jobs by combining both training throughput and memory constraints in the parameter server architecture as below.

\noindent\textbf{Throughput Modeling.}  
%We analyze the computation and communication patterns for DLRM training jobs the parameter server architecture.
The throughput of a DLRM training job represents the number of samples processed per unit of time. To model and predict the throughput, we divide one iteration time $T_{iter}$ into two parts: 
computation time \(T_{comp}\) and communication time \(T_{comm}\). Let $w$ denote the number of workers. Each worker consumes a mini-batch of data with size $m$ per iteration. Then, we formally model the throughput, denoted as $\Psi_{thp}$, as follows:

\vspace{-0.5em}

% \begin{equation}
% V = \frac{w \cdot m}{T_{lookup} + T_{grad} + T_{upd} + T_{sync}}\label{eq:throughput}
% \end{equation} 
% \begin{equation}
% \Psi_{thp}\left(w,p,\lambda_{w},\lambda_{p}\right) = \frac{w \cdot m}{T_{comp} + T_{comm}}\label{eq:throughput}
% \end{equation} 
\begin{equation}
\Psi_{thp} = \frac{w \cdot m}{T_{comp} + T_{comm}}\label{eq:throughput}
\end{equation} 
Note that the batch size $m$ remains unchanged during training.

\noindent\textbf{Computation Time Modeling.} The computation time \(T_{comp}\) includes two parts: the workers compute gradients ($T_{grad}$) and then the PSes update parameters using corresponding gradients ($T_{upd}$).

In each iteration, the gradient computation workload is proportional to the number of samples processed ($m$). The gradient computation rate is proportional to the number of parallel computing CPU cores in each worker (\(\lambda_{w}\)). Since the gradient update time (i.e., $T_{grad}$) can be calculated as the  workload divided by the rate, we formulate $T_{grad}$ as:

\vspace{-0.5em}
\begin{align}
 T_{grad}\left(\alpha_{grad},\beta_{grad}\right) = \alpha_{grad} \cdot \frac{m}{\lambda_{w}} + \beta_{grad}
\end{align}
where $\alpha_{grad}$ and $\beta_{grad}$ are learnable parameters representing how $T_{grad}$ scales with $m$ and $\lambda_{w}$ linearly. 

The parameter updating workload is proportional to the number of workers ($w$) as each worker computes one copy of the gradient and submits it to the PSes. The workload is also inversely proportional to the number of PSes ($p$) as all the PSes share these gradients. The update rate is proportional to the number of parallel computing CPU cores in each PS ($\lambda_{p}$). Therefore, we formulate $T_{upd}$ as follows:

\vspace{-1.5em}
\begin{align}
T_{upd}\left(\alpha_{upd},\beta_{upd}\right) = \alpha_{upd} \cdot \frac{w}{p\cdot\lambda_{p}} + \beta_{upd}
\end{align} 
where $\alpha_{upd}$ and $\beta_{upd}$ are learnable parameters representing how $T_{upd}$ scales with $p$ and $\lambda_{p}$ linearly. 

% In a DLRM training job, the computation time for a training step is primarily determined by two sequential phases: gradient generation and model parameter updates. 

% \textit{Modeling $T_{comp}$}. The workload for gradient generation exhibits a direct proportionality to the mini-batch size ($m$) and an inverse relationship with the CPU time slice allocation ($\lambda_{w}$). Thus, we represent the time for gradient computation, $T_{grad}$, as follows:
% \begin{align}
% T_{grad}(m,\lambda_{w}) =\alpha_{grad} \cdot \frac{m}{\lambda_{w}} + \beta_{grad}
% \end{align}
% where $\alpha_{grad}$, $\beta_{grad}$ are fittable parameters.

% \textit{Modeling $T_{upd}$}. After obtaining gradients, parameter servers update model parameters using optimization algorithms, such as Stochastic Gradient Descent (SGD). Over time, the size of incoming gradients scales linearly with the number of workers ($w$). The runtime has an inverse relation with the CPU time slice allocation ($\lambda_{p}$) to the parameter servers. Thus, we model $T_{upd}$ as follows:
% \begin{align}
% T_{upd}(p,\lambda_{p}) =\alpha_{upd} \cdot \frac{w}{p\cdot\lambda_{p}} + \beta_{upd}
% \end{align}

% where $\alpha_{upd}$, $\beta_{upd}$ are fittable parameters. 

\noindent\textbf{Communication Time Modeling.} The communication time \(T_{comm}\) includes two parts: 1) Workers pull parameters from PSes and push gradients to PSes to synchronize parameters (i.e., $T_{sync}$); 2) Workers lookup embeddings from PSes for gradient computation (i.e., $T_{emb}$).

For parameter synchronization, the network traffic (the amount of network communication data) between workers and PSes is twice the size of the model parameters ($M$) because both pulling and pushing operations transfer one copy of the data with size $M$. The network bandwidth $B$ is shared by \(w\) workers. The pushing-and-pulling workload is divided by the number of PSes $p$ as the model parameters and gradients are distributed across all PSes. Therefore, we formulate $T_{sync}$ as: 
% \begin{align}
% T_{sync}\left(w,p\right) = \alpha_{sync} \cdot \frac{w\cdot M}{p \cdot B} + \beta_{sync}
% \end{align}

\vspace{-2em}
\begin{align}
T_{sync}\left(w,p\right) = \alpha_{sync} \cdot \frac{M / p}{B / w} + \beta_{sync}
\end{align}
where $\alpha_{sync}$ and $\beta_{sync}$ are learnable parameters representing how $T_{sync}$ scales with $w$ and $p$ linearly. Here,  the model size $M$ and network bandwidth $B$ are considered constants during a short time.
\vspace{-0.5em}
% Two constants $M$ and $B$ are merged into $\alpha_{sync}$.

For embedding lookups, the network traffic is proportional to the samples \(m\) and the dimensions of the embedding table $D$, and it is shared by all PSes as the embedding table are distributed across PSes.
Therefore, we formulate $T_{emb}$ as: 

\vspace{-2em}
\begin{align}
T_{emb}\left(\alpha_{emb},\beta_{emb}\right) = \alpha_{emb} \cdot \frac{m \cdot D}{p} + \beta_{emb}
\label{eq:emb}
\end{align} 
where $\alpha_{emb}$ and $\beta_{emb}$ are learnable parameters representing how $T_{emb}$ scales with $m$ and $p$ linearly. Here, the embedding table dimension $D$ is fixed (hence being a constant) in the initialization of the embedding table.

Finally, we formally model throughput $\Psi_{thp}$ with a function $\mathcal{F}$ represented by the tuple of learnable parameters as follows:

\vspace{-2em}
\begin{align}
\Psi_{thp} = \mathcal{F}\left( \alpha_{grad},\beta_{grad},\alpha_{upd},\beta_{upd},\alpha_{emb},\beta_{emb},\alpha_{sync},\beta_{sync} \right) \label{eq: function}
\end{align}
\color{black}
\vspace{-2em}
\subsection{Optimization Formulation}
\label{subsec:optimization_formulation}
Given the resource-performance model in Eqn.~\ref{eq: function}, we formulate our optimization objective based on the \textit{"Resource Cost"} (i.e., for additional allocated resources) and the \textit{"Throughput Gain"} (i.e., from additional allocated resources) when scaling DLRM training jobs. \color{black} The goal of our optimization is to minimize the \textit{"Resource Cost"} while maximizing the \textit{"Throughput Gain"}.

%We define the resource allocation set (a.k.a allocation vector) based on the throughput model as previously discussed. 

\myline{Resource Cost Function (RC).} 
The resource scaling set, denoted as \(A\), represents the additional allocated resources to speed up a training job (e.g., the number of CPUs). Each type of resource in the set is denoted as $a_{r}$
(i.e., $A=\{a_{0}, a_{1}, ..., a_{r} \}$). Let $Money(a_{r})$ denote the expense a user should spend to allocate resource $a_{r}$. The \textit{"Resource Cost"} can be formulated as the sum of all resources' expenses (e.g., CPU and memory) :
\begin{align} 
RC(A) \ = \sum_{a_{r} \in A} a_{r} \times Money(a_{r}) \label{eq: rc}
\end{align}

\myline{Throughput Gain Function (TG).} We denote \textit{"Throughput Gain"} as the increased throughput that benefits from additional allocated resources. To illustrate, if we add 2 CPUs to a worker, the throughput might increase by 10 samples per second, namely the \textit{throughput gain}. However, increasing workers' resources, especially in cloud environments, practically comes with overheads (e.g., the time required to start a new worker equipped with 32 CPUs and 128GB of memory). Therefore, we formulate the \textit{Throughput Gain} as:
\begin{align}
    TG(A) = \ \Delta \Psi_{thp} - Overhead(A) \label{eq: tg}
\end{align}
Here, \(\Delta \Psi_{thp}\) represents the ideal increase in throughput if we neglect scaling overheads. \(Overhead(A)\) is the wasted training time caused by scaling the job with \(A\). This is estimated through statistical analysis based on the resource information of historical jobs within the cluster (e.g., the time required to start a worker/PS).

\myline{Multi-Objective Optimization.} Given the two functions Eqn.~\ref{eq: rc} \& \ref{eq: tg}, our goal is to find an optimal resource allocation set $A$, which minimizes the \textit{"Resource Cost"} while maximizing the \textit{"Throughput Gain"}. We formulate the optimization problem as follows:
\begin{equation}
\text{Objective:} \quad \quad \arg\min_{A}  (\, RC(A) \; , \; \frac{1}{TG(A)} \,) \label{eq:object}
\end{equation}
% We denote the benefit obtained from allocating resources \(x\) to a specific job as \(Benefit(x)\). This benefit is a function of the gain we get from the allocation, normalized by its associated cost, mathematically represented as:
% \begin{equation}
% Benefit(x) = \frac{Gain(x)}{Cost(x)} \ \label{eq:equal-c}
% \end{equation}
% Taking into account all the jobs in our training cluster, our goal is to maximize the aggregate benefits derived from various resource allocations. This can be formulated as:
% \begin{equation}
% \text{Objective:} \quad \arg\max_{x} \sum_{x \in X} Benefit(x)
% \end{equation}
% where \(X\) represents the collective resource allocation set for all training jobs in the cloud.

Given that Equ.\ref{eq:object} is neither linear nor convex, the problem cannot be solved using linear programming or convex optimization techniques and is NP-hard in general. To address this, we develop a heuristic \textit{auto-scaling} algorithm (see \S \ref{subsec:3-stage_auto-scaling_optimization three-stage-algorithm}).

\vspace{-1em}
\subsection{3-Stage Auto-Scaling Optimization}
\label{subsec:3-stage_auto-scaling_optimization three-stage-algorithm}
% Our importance design is motivated by multiple factors:
% 1.

% To solve the MOOP problem defined before, we first utilize an evolutionary algorithm (e.g., NSGA-II \cite{NSGA-II}) to generate a set of possible solutions  (e.g., a Pareto Frontier) for each job. After that,
% we select an optimal optimization plan according to the resources constraint  
% Given the complexity of the optimization objectives in the  MOOP and the huge search space created by the numerous jobs and myriad resource allocation options, finding an optimal solution for the problem is inherently challenging.
%Intuitively, considering the complexity of the optimization objectives of MOOP and the exponentially large search space (due to the number of jobs and myriad resource allocation options), it is hard to find an optimal solution for the problem. 
% Intuitively, to achieve the optimization objective in Eqn.~\ref{eq:object}, we design a scaling stage to generate  

\noindent\textbf{Intuition.} Given a DLRM training job, we first profile its runtime information for fitting the resource-performance model (Eqn.~\ref{eq: function}). Based on the model, \system generates an optimal resource plan with an \textit{auto-scaling} algorithm, aiming to achieve the optimization objective (Eqn.~\ref{eq:object}) (i.e., Scaling Stage \blacknumber{2} in Fig.~\ref{fig:system_dataflow}). 

To make the job training more robust in practice, we design a pre-scaling stage (stage \blacknumber{1}) to \textit{warm-starting} the job and a post-scaling stage (stage \blacknumber{3}) to handle the cloud instability. Specifically, compared to scaling the training job from scratch (i.e., cold start), users hope to see the submitted job performing well upon submission rather than waiting through a prolonged scaling process. Thus, we introduce a pre-scaling stage to allocate suitable start-up configurations. On the other hand, even provided with optimal resources, training jobs still encounter performance degradation (e.g., stragglers) due to cloud instability. Consequently, we introduce a post-scaling stage to ensure smooth training in the cloud.

\vspace{-1em}

\normalem
\begin{algorithm}[h]
    \SetAlgoLined
    \caption{Warm-Starting}
    \label{algo:data-driven-resource-setup}
    \SetKwInput{Input}{Input}
    \SetKwInput{Output}{Output}
    \Input{Historical Configurations $\mathcal{D}$, New Job $\mathcal{J}$, Exponential Smoothing Function $\mathscr{E}$, \quad \quad \quad Smoothing Factor $\mu$ $(0 < \mu < 1)$}
    \Output{Warm-Starting Resource Allocation $\bar{A}^{k-1}$}

    \BlankLine
    \SetKwInput{Topk}{Identify Top-K Similar Jobs of $\mathcal{J}$ With MetaData}
    \SetKwInput{Smoothing}{Initialize Smoothing for Configuration}
    
    \Topk{}
    $\{A^{0}, A^{1}, ..., A^{k-1}\}$ $\gets$ top-$k$ similar job configuration in $\mathcal{D}$;

    \BlankLine
    \Smoothing{}
    Rank $\{A^{0}, A^{1}, ..., A^{k-1}\}$ with similarity; \\
    Initialize smoothed configuration: $\bar{A}^{0} = A^{0}$; \\
    
    % \For{$i$ = $1$ to $k$}{
    %     \For{ $a^{i}_{r}$ \in $A^{i}$}{
    %         Apply $\mathcal{E}$ to update the smoothed configuration       $\bar{a}^{i}_{r}$:\\
    %         $\bar{a}^{i}_{r} \gets \mu \times a^{i}_{r} + (1 - \mu)   \times \bar{a}^{i-1}_{r}$;
    %     }
    % }
    \For{$i$ = $1$ to $k - 1$}{
        Apply $\mathscr{E}$ to get the smoothed configuration   $\bar{A}^{i}$:\\
        $\bar{A}^{i} \gets \mu \times A^{i} + (1 - \mu)   \times \bar{A}^{i-1}$;
    }

    \BlankLine
    
    \Return $\bar{A}^{k-1}$\;
\end{algorithm}
\ULforem

\vspace{-1em}

\noindent\textbf{\blacknumber{1} Pre-scaling Stage: Warm-Starting.} 
\label{sec:start-up-stage}

As shown in Algorithm \ref{algo:data-driven-resource-setup}, we adopt a \textit{warm-starting} algorithm to identify a suitable start-up resource configuration. We first use the job's features (e.g., model metadata) to collect top-k similar jobs. Specifically, given historical job configurations stored in the \textit{Configuration Database} $\mathcal{D}$, the algorithm first calculates the similarity in each type of feature and then gets the top-k similar job configurations where the configuration of \(i\)-th similar job is denoted as $A^{i}$ ($A^{k-1}$ is the job configuration with highest similarity). We first initialize the target configuration set as $\bar{A}^{0} = A^{0}$. Subsequently, we use the \textit{Exponential Smoothing Function} $\mathscr{E}$ to generate the smoothed configuration $\bar{A}^{i}$ in an iterative manner. Formally, for each $A^{i}$, we calculate $\bar{A}^{i}$ as follows:
\begin{align}
\mathscr{E}: \bar{A}^{i} = \mu \times A^{i} + (1 - \mu)   \times \bar{A}^{i-1}
\end{align}
 where the smoothing factor $\mu$ balances the influence of historical configurations, determining the weight between the job's configuration $A^{i}$ and the result of last iteration $\bar{A}^{i-1}$. Lastly, we use the final iteration result $\bar{A}^{k-1}$ as the start-up job configuration.

\noindent\textbf{\blacknumber{2} Scaling Stage: Auto-Scaling.}
\label{sec:auto-scaling-stage}

We aim to auto-scale training jobs in this stage according to our resource-performance model (Eqn.~\ref{eq: function}). Auto-scaling includes three steps: 1) online model fitting, 2) job-level resource plan candidate generation, and 3) cluster-level weighted greedy selection.

\noindent\textbf{Online Model Fitting.} As detailed in \S \ref{subsec:resource-performance_modeling}, we represent the throughput of a training job by a group of $\alpha$ and $\beta$ parameters (e.g., $\alpha_{grad}$, $\beta_{grad}$). To build our resource-performance model, these parameters can be fitted based on runtime profiles. \system continuously monitors the time taken for each iteration $T_{iter}$ to measure the throughput $\Psi_{thp}$ as in Eqn.\ref{eq:throughput}. At a fixed interval,~\system refines the groups of $\alpha$ and $\beta$ using the accumulated data, minimizing the root mean squared logarithmic error (RMSLE) between the theoretical model and the actual data (by employing Non-Negative Least Squares (NNLS)\cite{NNLS}). Note that all parameters $( \alpha, \beta)$ are bound to remain non-negative.

\noindent\textbf{Job-Level Resource Plan Candidates Generation.} After model fitting, we learn the function $\mathcal{F}$ as Eqn.~\ref{eq: function} representing the relation between resource allocation $A$ and throughput $\Psi_{thp}$. We utilize NSGA-II~\cite{NSGA-II} to generate resource allocation plans that meet the Pareto Frontier. The Pareto Frontier represents the set of all optimal allocations that cannot be improved on one dimension without worsening another. For example, we can not increase \textit{"Throughput Gain"} without increasing the \textit{"Resource Cost"}. NSGA-II is an evolutionary algorithm known for its rapid convergence to the Pareto Frontier in low-dimensional multi-objective problems. 

% \textcolor{red}{Subsequently, we select an optimization plan from these candidates using \textit{weighted greedy selection}}.

\noindent\textbf{Cluster-Level Weighted Greedy Selection.}
% In this stage, the aggregated execution plans at the cluster level are sourced from the Job Master. These plans undergo sorting based on their resource efficiency. The culmination of this stage is to transition the job configuration \( X \) to an optimized variant \( X' \) relying on these methodically ordered plans.
With all the optimization plan candidates for each job, we employ \textit{weighted greedy selection} to determine the final execution plan for each job. We denote the \textit{"Resource Efficiency"} for allocating resources \(A^{j}\) to a specific job $j$ as \(RE(A^{j})\). \(RE(A^{j})\) is a function of the \textit{"Throughput Gain"} we get from the allocation, normalized by its associated \textit{"Resource Cost"}, mathematically represented as:
\begin{equation}
RE(A^{j}) = \frac{TG(A^{j})}{RC(A^{j})} \ \label{eq:equal-c}
\end{equation}

\vspace{-0.5em}

To determine a set of efficient cluster-wide resource allocations, we maximize the weighted benefit sum of each job:

\vspace{-0.5em}

\begin{align}
& \text{Weighted Greedy:} \quad \arg\max_{A^{J}} \sum_{j \in J} RE(A^{j}) \cdot WG(A^{j})\\
& \hspace{8em} \text{subject to:} \sum_{j \in J} A^{j} \leq S
\end{align} 

Here, $J$ represents the jobs needing reallocation; $S$ denotes the total resources; $WG(A^{j})$ denotes the priority value determined by a range of priority algorithms tailored to the cluster's preference. In our cluster, we use the remaining time for each training job (represented by the remaining samples divided by the throughput) to calculate the weight $WG(A^{j})$ as follows:

\vspace{-1.5em}
\begin{align}
 WG(A^{j}) = \frac{1}{\left(\Phi^{j}_{sp}/\Psi_{thp}^{A^{j}} + \epsilon\right)^{\rho}}
\end{align}

Here, $\Phi^{j}_{sp}$ represents the number of the remaining samples to be trained in the job $j$. As $\rho\rightarrow0$, $WG(A^{j})$ smoothly approaches $1$ for every job $j$, which means we consider the weights of all jobs to be equal. When $\rho \rightarrow \infty$, $WG(A^{j})$ can prioritize jobs with a shorter completion time. In contrast, as $\rho \rightarrow -\infty$, $WG(A^{j})$ prioritize jobs with longer completion time. A cluster operator may select a suitable value for $\rho$, based on practical priorities. At \company, we choose $\rho = 2.5$ as a reasonable value to complete shorter jobs quicker and release the resources. Additionally, $\epsilon$ denotes a very small value used to prevent division by zero. \color{black}

\noindent{\bf Plug-in Algorithm API.} Although the auto-scaling algorithm based on \textit{weighted greedy} is sufficiently efficient in the settings of \company (see \S \ref{subsec:production_adoption_evaluation}), this algorithm may diverge from reality for specialized hardware. Hence, \system provides a standard API to ensure other customized algorithms can be plugged in easily.

\noindent\textbf{\blacknumber{3} Post-scaling Stage: Instability Handling.}
\label{subsec: stage-3}

During the auto-scaling phase (stage \blacknumber{2}), we assume that the job is interference-free. Yet, in practical cloud environments, this assumption might not hold as discussed in \S \ref{subsec:challenges_of_DLRM_training_at_antgroup}. Therefore, \system involes many techniques to handle various cloud instabilities in the post-scaling stage (stage \blacknumber{3}). We highlight some key instability issues and the techniques adopted by \system to address them in the following (with more details in \S \ref{sec:handling_instability}). 

\noindent\textbf{Worker Stragglers.} In heterogeneous clusters, certain worker pods may be assigned to physical machines with slow hardware (i.e., low-frequency CPU and/or low-speed memory) or be hindered by high-priority pods due to resource contention. These worker stragglers can result in submitting stale model gradients to the parameter servers, leading to decreased model accuracy and longer training time~\cite{stale}. To address this, \system~implements a \textit{dynamic data sharding} mechanism to minimize the discrepancy in iteration rounds among workers (see \S \ref{subsec: dynamic_data_sharding}).

\noindent\textbf{PS Stragglers.}
Our DLRM jobs run on the TensorFlow framework, which determines parameter allocation based on \textit{tensors} or multi-dimensional arrays. The size of tensor-based parameters assigned to PSes can differ substantially, resulting in unbalanced workloads \cite{tensorflow2015-whitepaper}. Consequently, PSes performing large matrix multiplications with more allocated parameters experience significantly higher CPU loads, resulting in the PS stragglers. To address this, we adopt DeepRec~\cite{DeepRec}
%\footnote{https://github.com/DeepRec-AI/DeepRec} 
to ensure that the embedding parameters are evenly distributed across the new set of PS nodes. 

\noindent\textbf{Scaling Overhead.} As formulated in Eqn.~\ref{eq: tg}, scaling overhead plays an important role in workload migration. To speed up the migration process, \system introduces a \textit{seamless migration} mechanism to mitigate the scaling overhead (see \S \ref{subsec:seamless_migration}).

\noindent\textbf{Out of Memory Problem.} Uneven allocation and high memory consumption from large embedding tables can lead to out-of-memory problems in PSes. To address this, \system invents a \textit{OOM predition} mechanism (see \S \ref{subsec:oom_prevention}).

%In our DLRM jobs running on the TensorFlow framework, parameter allocation is determined at the granularity of \textit{Tensors} (i.e., a multi-dimension array), which can vary significantly in size. This variability can result in unbalanced workloads, particularly with larger matrix multiplications caused by larger-sized parameters, leading to hot parameter server issues. To mitigate this, \system monitors nodes for high CPU loads and introduces more capable parameter servers with additional resources for migration to alleviate bottlenecks. To speed up the migration process, \system implements a \textit{Seamless Migration} mechanism. Moreover, uneven allocation and high memory consumption from large embedding tables can lead to out-of-memory problems in parameter servers. To address this, \system introduces a \textit{Memory Pre-adjustment} mechanism (see \S \ref{subsec: seamless_migration_and_OOM_handling}).

% Conversely, if a node exhibits significant underutilization, \system adds extra nodes. We continuously monitor the CPU load and training speed, adjusting the system until the node's CPU load surpasses a threshold or additional nodes no longer enhance the training speed. 

% \textit{Hot Spot.}
% \textit{Out-of-Memory (OOM) Situations.}
% In sparse models, the number of items in the PS embedding table can escalate, causing the PS memory to expand. Should a node's memory usage approach a risk threshold, DLRover substitutes it with a more memory-rich node.
% \textit{Low CPU Utilization in PS.}

\vspace{-1.5em}
\section{Handling Instability}
% \section{Implementation}
\label{sec:handling_instability}
\system's design draws upon the collective insights and practices from the realm of distributed/training systems. In this section, we focus on the key mechanisms developed by \system to enhance the performance and reliability of the DLRM training system.

\vspace{-1.5em}
\subsection{Dynamic Data Sharding}
\label{subsec: dynamic_data_sharding}
\color{black}\system introduces a \textit{dynamic data sharding} mechanism that enables fine-grained data serving by partitioning training data into \textit{numerous small shards of various sizes}. These data shards can be on-demand, dynamically assigned/reassigned to 1) slow workers (i.e., stragglers) to balance their paces of data processing and model updates with their peers for consistent model quality; and 2) new/healthy workers for fast elasticity or fault tolerance.
%`\Hao{discussion}. 
%and indexing every data shard. 
%Based on this mechanism, \system handles imbalanced workloads caused by worker stragglers and lost data caused by pod failure.
\begin{figure}[h]
\vspace{-1em}
\centering
\includegraphics[width=\columnwidth]{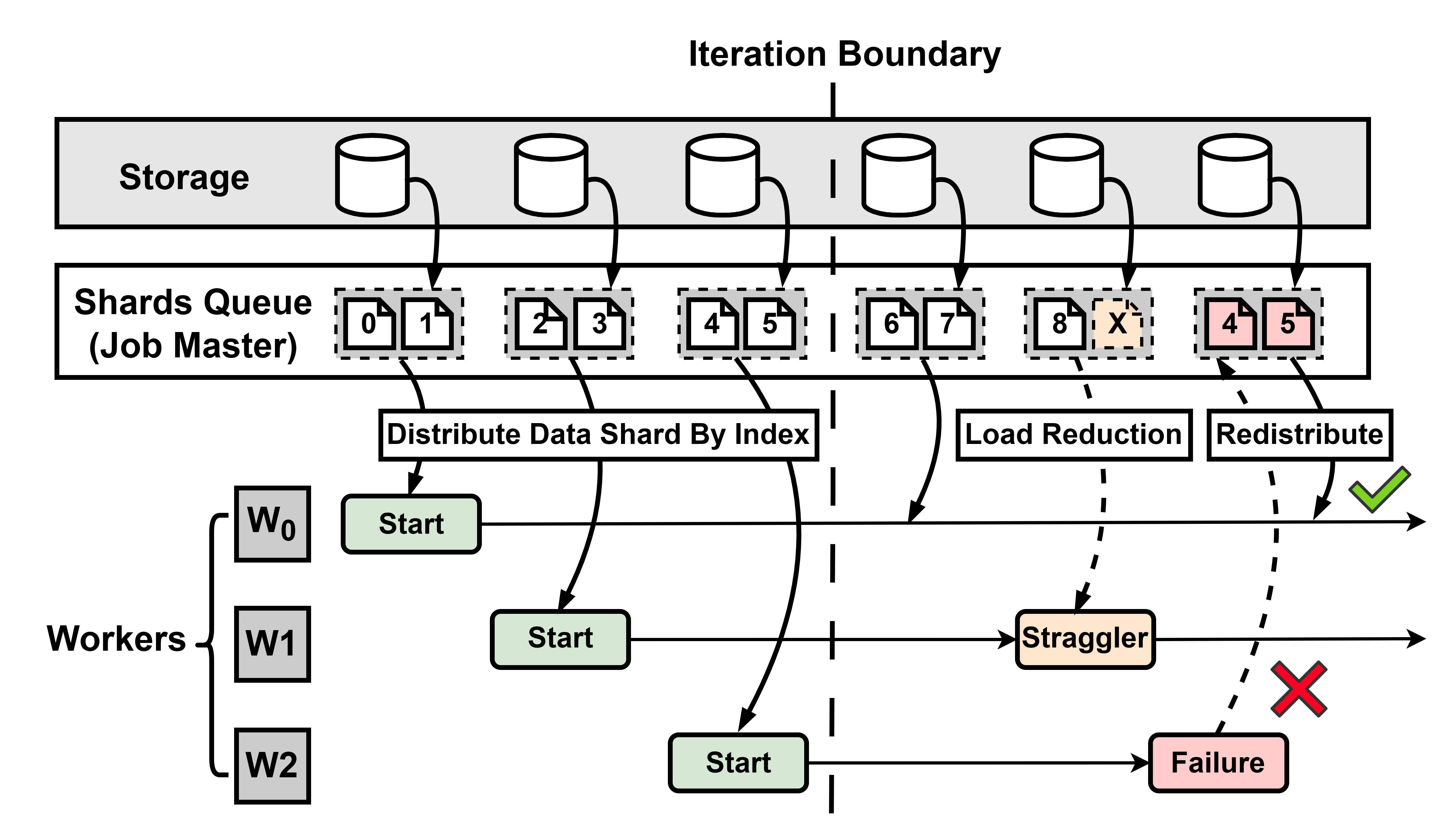}
\vspace{-2.2em}
\caption{Dynamic data sharding.}
\label{fig:dynamic_data_shard}
\vspace{-2.5em}
\end{figure}

\noindent\textbf{Data Sharding of \system.} 
Traditionally, training data are (statically) partitioned and distributed among workers at the beginning of job training. In contrast, \system splits the dataset into \textit{numerous, much smaller, and variably-sized} shards (e.g., 64, 128, or 256 data batches), each labeled with a unique index; \system then manages the data serving by delivering such fine-grained data shards to the corresponding worker \textit{on-demand} during a worker's life cycle.
%In Fig.~\ref{fig:dynamic_data_shard}, unlike traditional data partitioning which only distributes training data between workers at the beginning of training jobs, \system splits the dataset into \textit{numerous, much smaller} shards (e.g., 4 data batches), each labeled with an index. \system manages the data serving dynamically by delivering on-demand data shards to the corresponding worker during the worker's life cycle. 

Specifically, as illustrated in Fig.~\ref{fig:dynamic_data_shard}, data shards are organized within a \textit{shards queue}. Upon initiating job training, a worker fetches the required data shard (by its index) from this queue. Note that workers are initially assigned comparable workloads, as determined by the quantity of data shards and the number of data samples within them over a fixed time interval. Throughout the training process, workers dispatch heartbeat packets at regular intervals to the job master (Fig.~\ref{fig:system_dataflow}), which include the number of data samples they have processed, named \textit{progress offset}. These heartbeat packets and the progress offsets serve three critical functions: 1) They signal that the worker is operational and active. Conversely, this mechanism is employed to identify workers that have failed. 2) The job master uses the \textit{progress offset} to gauge the worker's training progress and determine if it is falling behind -- i.e., identifying any straggler if the offset is noticeably lesser than its peers. 3) Upon completion of the designated data shard, the worker sends its final \textit{progress offset} to the job master, indicating the completion of that shard's processing. Subsequently, the worker acquires a new shard from the shards queue to proceed with the training.

%Partitioning data into smaller data shards and adopting the \texttt{shards queue} gives rise to the benefit of enabling fast elasticity while ensuring consistent model without re-partitioning the data among workers.

%and offset help \system scale without re-partition the data between workers when a worker is going to join the worker group. For example, the new worker can retrieve a data shard easily from the {shards} queue without interfering with other workers.

%We prepare these data shards in a \texttt{shards} queue. When a worker starts operating, it retrieves a data shard using the data shard index it should consume. The worker uses the offset of processed data samples (e.g., the number of processed data samples) as a heartbeat package to inform the job master that it is alive. Meanwhile, the job master uses the offset to monitor the training status of the worker and identify whether the worker is a straggler (e.g., a slow worker). Once the offset reaches the end of the data shard, the job master marks this data shard as \texttt{done}. Then, the worker retrieves a new data shard and continues training. The \texttt{shards} queue and offset help \system scale without re-partition the data between workers when a worker is going to join the worker group. For example, the new worker can retrieve a data shard easily from the {shards} queue without interfering with other workers.
\noindent\textbf{Failure Recovery.} 
Once the job master does not receive the heartbeat package from a worker for a reasonably long time, it is identified as a failure event. In the event of a worker failure (highlighted in red in Fig.~\ref{fig:dynamic_data_shard}), the job master re-joins the unfinished data shard(s) of the failed worker to the \textit{shards queue}, awaiting redistribution to another healthy worker. This mechanism simply ensures that the training job consumes the training data \textit{without} any data omission or duplication, guaranteeing the model's consistent quality.

\noindent\textbf{Handling Stragglers.} 
Throughout job training, the job master also keeps track of the \textit{progress offsets} each worker provides. A worker is labeled a straggler if it lags significantly behind its peers (highlighted in orange in Fig.~\ref{fig:dynamic_data_shard}). In such cases, the system mitigates the issue by reassigning the slower worker a smaller workload, such as providing a data shard with fewer batches (e.g., scaling down from a shard with 256 batches to one with 128). This way, the system can dynamically tailor the volume of data samples a slower worker has to process before it submits its gradients to the PSes. Consequently, it enables the straggling worker to align its gradient submission rate with others, preventing the submission of stale gradients and maintaining consistent model quality.
%from potential conflicts between outdated gradients from the straggler and newer gradients from other workers. 
%\system monitors the offset of processed data samples reported by the workers and evaluates the processing speed.  If a worker is too slow than other workers, it is identified as a straggler (highlighted in orange in Fig.~\ref{fig:dynamic_data_shard}), \system adjusts the workload by assigning a data shard with fewer samples to this worker (e.g., from a data shard with 4 batches to a data shard with 2 batches). This action adjusts how many samples workers need to process before submitting the gradients to the PSes. Thus, we ensure the straggler can catch up with the gradient submission speed of other workers, preventing the submission of excessively outdated gradients. This protects the model quality from potential conflicts between outdated gradients from the straggler and newer gradients from other workers. 

\noindent\textbf{Fast Elasticity.} Partitioning data into smaller data shards and adopting the \textit{shards queue} ultimately gives rise to the benefit of enabling fast elasticity because any new worker -- i.e., after a stop-and-restart with adjusted CPU/memory resources -- can simply retrieve a data shard from the \textit{shards queue} without the process of data re-partitioning and redistribution among all workers. 
\vspace{-0.1in}
%\system keeps monitoring the heartbeat package of the workers. Once \system can not receive the heartbeat package for a long time, it is identified as a failure event. In the event of a worker failure (highlighted in red in Fig.~\ref{fig:dynamic_data_shard}), the job master re-join the unfinished data shard of the failed worker to the \texttt{sharding} queue, awaiting redistribution to a healthy worker. This mechanism ensures the training job consumes the training data without omission or duplication, guaranteeing the model's reproducibility

\subsection{Seamless Migration}
\label{subsec:seamless_migration}
\system develops a \textit{seamless migration} mechanism to minimize resource scaling overhead during training \textit{by strategically overlapping the scaling progress with ongoing training activities}. 
This approach effectively reduces training downtime associated with worker/PS initialization and delays in allocating new resources. %This overlapping saves efficient training time from workers/PSes initialization and the pending time for requesting new resources. 
To further reduce the scaling overhead, 
\system invents a \textit{flash-checkpoint} mechanism, 
%that enables fast checkpointing 
%40 GBs model in seconds to reduce wasted training time on checkpoints by writing checkpoints to 
which accelerates checkpointing through the use of shared memory and asynchronous data persistence.
%and persisting the checkpoints to disk asynchronously.

\begin{figure}[t]
\label{fig:sync_checkpoint}
\centering
\includegraphics[width=\columnwidth]{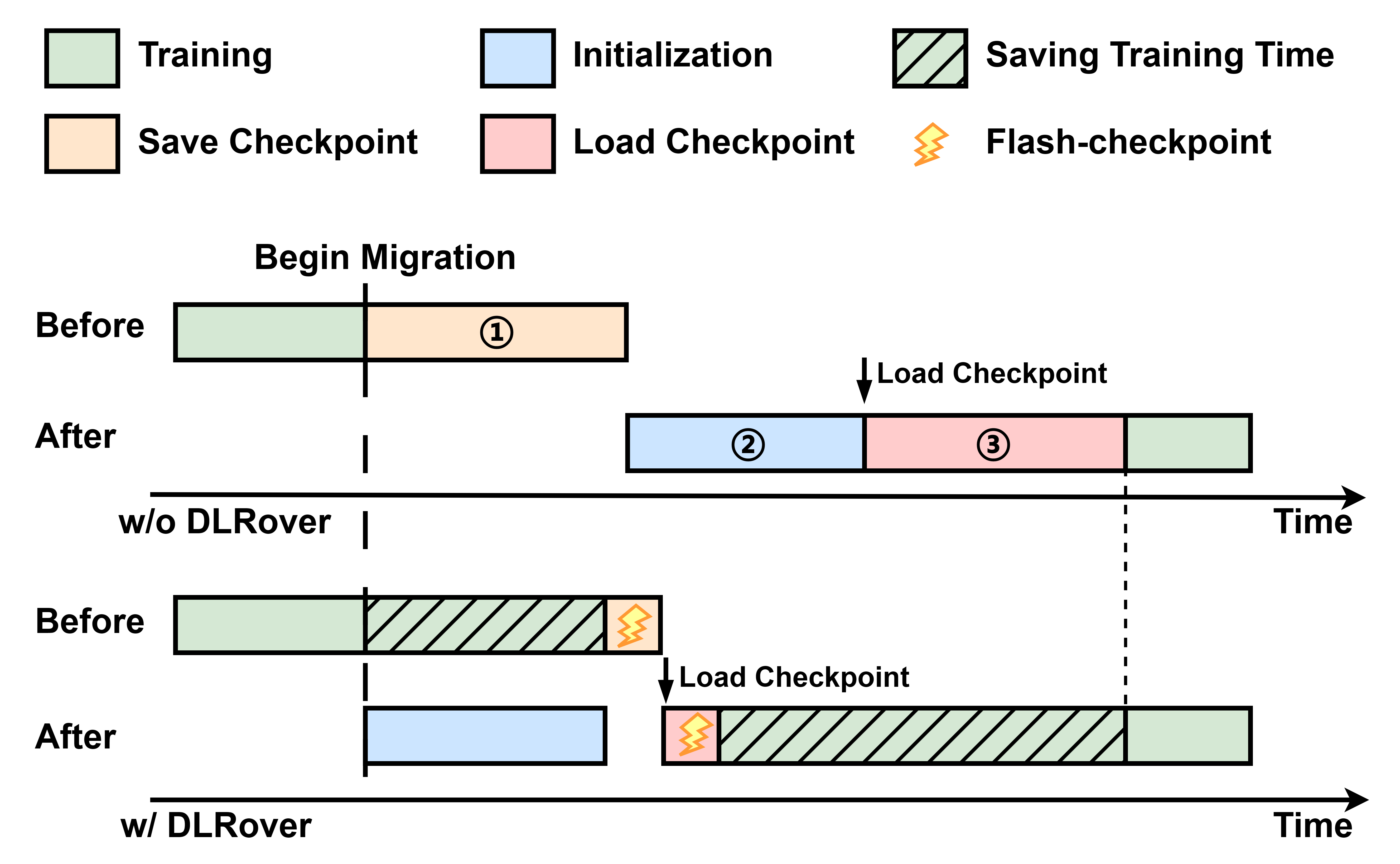}
\vspace{-3em}
\caption{Seamless Migration.}
\label{fig: seamless_migration}
\vspace{-2.2em}
\end{figure}

\label{subsec:seamless_migration_and_OOM_handling}
\noindent\textbf{Seamless Migration.} Scaling resources up/down conventionally involves a \textit{stop-and-restart} operation: As shown in Fig.~\ref{fig: seamless_migration}, the migration progress (w/o \system \footnote{In this work, "DLRover" and "DLRover-RM" are used interchangeably.}) \ding{172} stops the workers/PSes and checkpoints the model parameters in (remote) persistent storage (e.g., the RDS at \company); \ding{173} deploys and initializes new PSes and workers based on the updated resource plans -- including adjusting the number of worker/PS pods and/or their allocated resources, pulling images of worker/PS pods, and launching the new worker/PS pods; and \ding{174} finally loads the checkpoints from the persistent storage and resumes the training job. While being straightforward, the \textit{stop-and-restart} operation results in nontrivial pending time (e.g., up to tens of minutes), especially in the cloud environment (\S \ref{subsec:challenges_of_DLRM_training_at_antgroup}).
% as discussed in Section~\ref{subsec:challenges_of_DLRM_training_at_antgroup}.

%Conventionally, when the number of workers or parameter servers assigned to a job changes, the training framework checkpoints the model parameters and saves them to (remote) persistent storage. Then, the traditional training framework stops the workers/PSes in the old job and restarts the job from the checkpoint. Then redeploy parameter servers and workers based on the new resource plans. This process wastes efficient training time on the initialization progress of a new job. With more details, as depicted in Fig.~\ref{fig: seamless_migration}, traditional migration progress (w/o \system) includes saving checkpoints first then initializing new resource pods, and finally loading checkpoints. Typically, the initialization of the new job includes requesting resources for new pods, pulling images, and launching the jobs. Notably, the \textit{stop-and-restart} approach may encounter nontrivial pending time during requesting resources for new pods, especially in the cloud environment. For example, the widely used cloud management platform Kubernetes processes resource requests based on priority while training jobs are often considered low priority to request the resource. 

We observe that such a synchronous \textit{stop-and-restart} operation can be decoupled into two parallel ones. Initially, rather than beginning with step \ding{172}, \system starts initializing and deploying new workers and PSes (i.e., \ding{173}) while allowing the "old" workers/PSes to proceed with the ongoing training job. Once all the new workers/PSes are ready to use, \system performs steps \ding{172} and \ding{174}, during which training jobs must be paused. To further speed up the critical path (\ding{172} and \ding{174}), \system invents the \textit{flash-checkpoint} approach.

\noindent\textbf{Flash-checkpoint.} 
%eeds to hang up the training process for a few minutes. 
In the production environment of \company, the RDS services are commonly shared among various internal services, each allocated a limited bandwidth. This, unfortunately, prolongs the critical path of DLRover-RM's migration (\ding{172} and \ding{174}). Unlike checkpointing for fault tolerance, which focuses on data persistence, checkpointing during the migration (of workers and PSes in \system) should emphasize speed over strong persistence. This insight motivates \system to leverage a distributed caching service (at AntGroup~\cite{alluxio}) to enhance migration efficiency, called \textit{flash-checkpoint}. 
%As depicted in \ref{fig: flash_checkpoint}, 
During migration, \system checkpoints (\ding{172}) or loads (\ding{174}) model parameters via the caching service. As the bandwidth and access speed of the caching service is much faster than the RDS, the checkpointing process takes much less time (e.g., less than 1 second for a 20GB model). Further, the caching service facilitates data sharing between new and old workers/PSes when they are located on the same physical node, eliminating network transmission.

Note that \system has a separate RDS-based checkpointing mechanism for job-level and system-level fault tolerance (i.e., to recover a job or the system from failures). The \textit{flash-checkpoint} mechanism, though separately, enhances the RDS-based checkpointing by flushing model parameters from the caching system to the RDS asynchronously (i.e., more update-to-date checkpoints).

%The low bandwidth of RDS severely increases the cost of checkpointing. To save wasted time on blocking and writing checkpoints to low-bandwidth storage, we implement an in-memory {flash-checkpoint} mechanism. To illustrate, \system leverages the spare memory in PSes nodes as a checkpoint cache to store checkpoints in seconds. 
%As illustrated in Fig.~\ref{fig: flash_checkpoint}, we design a \textit{two-loop} checkpointing scheme considering hierarchical storage with different bandwidths. In the \textit{training loop}, when \system starts to save checkpoints, \system first allocates a shared memory for holding the model parameters and then saves the model parameters to this pinned memory (this process takes less than 1s). After that, in the \textit{persistent loop}, the job master initiates a sub-process to asynchronously save model parameters to RDS minimizing the interference to the training process. \Lan{we just design the minimum fault tolerance, and more detailed implementation can refer to GEMINI}

%\begin{figure}[t]
%\label{fig:flash_checkpoint}
%\centering
%\includegraphics[width=0.8\columnwidth]{figures/flash_checkpoint.png}
%\vspace{-0.2in}
%\caption{Flash-Checkpoint.}
%\label{fig: flash_checkpoint}
%\vspace{-0.2in}
%\end{figure}

\vspace{-1em}

\subsection{OOM Prevention}
\label{subsec:oom_prevention}
%\color{blue}
\system employs a prediction mechanism for preventing out-of-memory (OOM) via modeling the dynamics of memory usage.
%\system involves an \textit{OOM-prediction} mechanism to prevent OOM errors by modeling the 
%increasing part 
%dynamics of memory consumption. 
%We first formulate the memory consumption $M_{emb}$, then \system uses $M_{emb}$ to predict memory usage and checks if PSes might exceed the capacity before the job completion (e.g., 10,000 steps to complete). If the prediction step is within the job completion step, \system scales the PSes with a higher memory capacity.

%\noindent{\bf Memory Pre-Adjustment.} 
Memory consumption during DLRM training consists of two main components: 1) a static portion, which includes model parameters, gradients, and optimizer states, and 2) a variable portion, such as the embedding table. As discussed in \S ~\ref{subsec:challenges_of_DLRM_training_at_antgroup}, the size of the embedding table can significantly expand as training advances. Our prediction efforts for potential out-of-memory (OOM) concentrate on the memory changes of the embedding table.
%Memory usage in DLRM training can be divided into two parts: 1) a constant part (e.g., model parameters, gradients, and optimizer states) and 2) a dynamic part (e.g., the embedding table). As discussed in \S ~\ref{subsec:challenges_of_DLRM_training_at_antgroup}, the size of the embedding table can increase dramatically during the training progress. We focus on the memory used by the embedding table to predict potential OOM problems. 

Let $D$ represent the embedding table dimension, $T$ the data type (e.g., float32), and $\phi_{cats}$ the number of categories in the embedding table. The embedding memory usage, $M_{emb}$, is thus given by $M_{emb} =  T \cdot D \cdot \phi_{cats}$, where 
$\phi_{cats}$ increases as new features integrate during training. We assume that $\Delta \phi_{cats}$ (i.e., changed categories) is proportional to the data consumption rate. Hence, the newly integrated categories are given by: $\Delta \phi_{cats} \propto \Psi_{thp} \cdot \Delta t$. 
This indicates that the memory change of the embedding table, $\Delta M_{emb}$, is proportional to $\Delta t$. Therefore, the memory increase can be estimated with the throughput profiles (see Eqn.~\ref{eq: function}).
After modeling $M_{emb}$, \system can use it to predict memory usage and check if PSes would exceed the memory capacity before the job completion (e.g., 10,000 steps to complete). If an OOM error is estimated to occur within the job completion step, \system scales the PSes with larger memory capacity.
\vspace{-1.5em}
\section{Evaluation}
%\vspace{-0.1in}
\label{sec:evaluation}

\noindent{\bf Production Adoption.}
\sloppy
We have implemented \system with approximately 20,000 lines of Python and 12,000 lines of Golang. \system is developed as a cloud-native service for Kubernetes and can be openly accessed at \textcolor{blue}{\url{https://github.com/intelligent-machine-learning/dlrover}}. \system has been widely deployed in \company's production environment, supporting 95\%+ of  \company's DLRM training jobs (2K per day) and utilizing 60K-120K CPU cores and 80-220TB memory per day. Additionaly, 10+ major tech companies have also adopted \system in their production environments. 
\sloppypar

\noindent{\bf Evaluation Environments.} We thoroughly evaluated \system using both a small-scale (and more controlled) cluster and \company's production environment. 1) The small-scale cloud-based cluster has 20 CPU servers, each equipped with two 16-core Intel Xeon E5-2682 @2.5GHz CPU and 192GB RAM. All experiments were conducted on a Kubernetes-managed cluster.
%with a maximum allocation of 200 CPU cores. 
%The cloud cluster is managed by Kubernetes.
%\noindent{\bf Production Environment.} 
%we can cite the cougar env to introduce this
% https://ieeexplore.ieee.org/abstract/document/10184667 
2) We further evaluated \system by deploying it in our production environment. The cloud-based cluster provides $\sim$1.6 million CPU cores, 3.24 PB of memory, and 344 PB of disk storage. Different types of jobs (e.g., training, serving, and stream processing) are %consolidated in the cloud and 
sharing these cloud-based resources. 

\noindent{\bf Workloads.} 
We employed three representative DLRM models at \company \cite{ant_model_wd1,ant_model_wd2,ant_model_dcn1,ant_model_wd1}: 
1) \textbf{Model-X}: Wide \& Deep \cite{wide-deep}; 
2) \textbf{Model-Y}: xDeepFM\cite{xdeepfm}, and 
3) \textbf{Model-Z}: DCN\cite{DCN}. 
We measured the performance of \system on Criteo dataset~\cite{website:criteo}. 
All models were implemented in TensorFlow 1.13, and each job had the same batch size of 512 with 200,000 training steps. 
%To show the effectiveness of {\sc DLRover}, we manually fine-tuned the "optimal" resources for jobs without the support of  \system 
%(i.e., w/o DLRover)  in the testbed experiment.

\noindent{\bf Comparison Baselines.} The evaluation covers two primary baselines: 1) w/o \system: the general distributed framework (e.g., Kubeflow) used in the cloud~\cite{kubeflow2023} without \system support. In this baseline, each job’s resources needed to be manually configured. To show the effectiveness of \system, we first well-tuned the resource configuration of this baseline and compared it with the support of \system (i.e., w/ {\system}).
2) The state-of-the-art deep learning training job schedulers for \textit{CPU-only} scenarios (i.e., as discussed in \S\ref{sec:background_and_motivation}, \system is designed for CPU-only hardware), including Elastic Scheduler (short as \textbf{ES}) \cite{mlsys-elastic} and \textbf{Optimus}~\cite{Optimus}. These schedulers are well-designed for traditional deep learning models in NLP and CV. Note that, both \textbf{ES} and \textbf{Optimus} add or remove a fixed number of nodes each time, while \textbf{ES} only modulates workers and \textbf{Optimus} adjusts PS or workers. 

% The resource adjustment interval was set to 3 minutes, encompassing 20 seconds for launching Kubernetes pods, 100 seconds for initializing training, and 60 seconds for sampling.  

\noindent{\bf Metrics.} We focus on four kinds of metrics. 1) \textit{Job Completion Time (JCT)}: the end-to-end model training time (the shorter is better); 2) \textit{Job Completion Rate (JCR)}: the proportion of successfully completed jobs to the total jobs submitted within a defined timeframe. A higher JCR indicates efficient job processing and a lower fault ratio; 3) \textit{CPU Utilization Rate (CUR)}: the workload processed by the CPU within a specified time period; and 4) \textit{Memory Utilization Rate (MUR):} the volume of system memory consumed during operations.

\vspace{-1.5em}
\subsection{End-to-End System Performance} 
\label{end-to-end_system_performance}

In this section, we demonstrate the end-to-end performance of \system within the small-scale cluster, which eliminates the impact of cloud instability and ensures a more fair comparison.

\begin{figure}[h]
\vspace{-1em}
\hspace{-1em}
\includegraphics[width=\columnwidth]{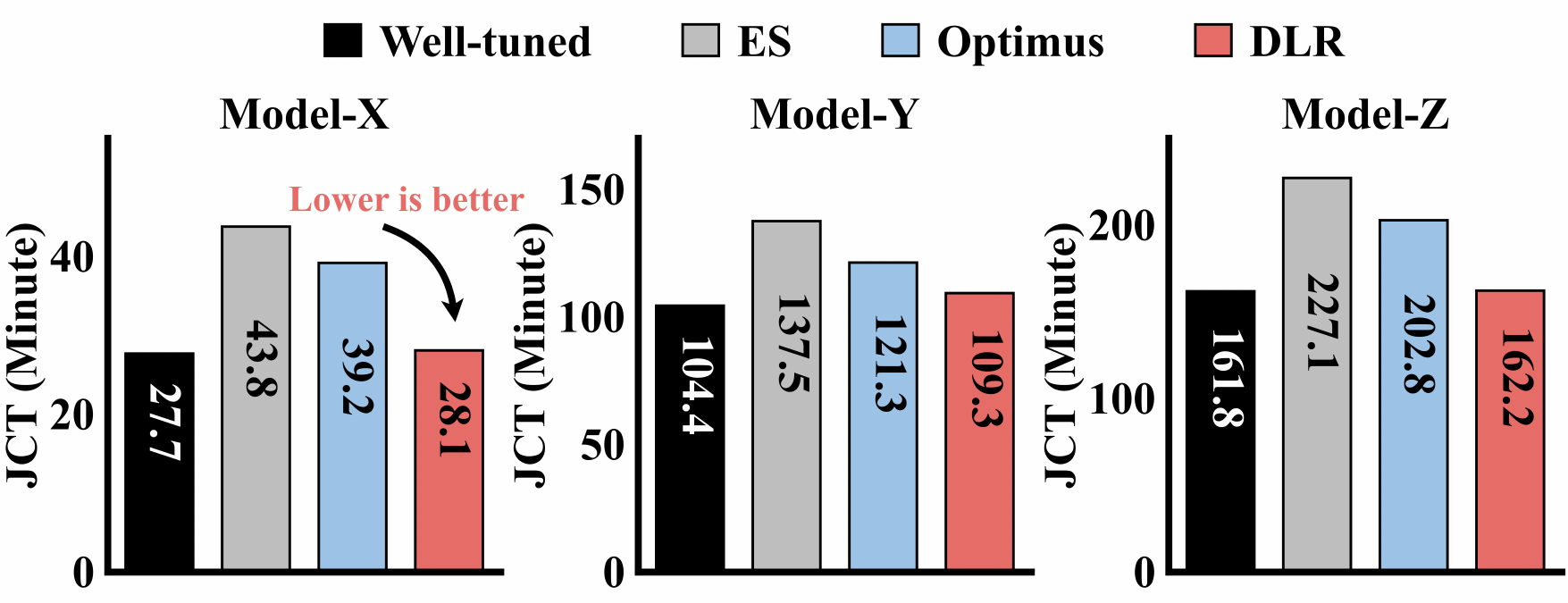}
\vspace{-1.5em}
\caption{DLRover-RM achieves comparable JCT to well-tuned
 configurations and reduces
 JCT compared with competitors.}
\vspace{-1em}
\label{fig:jct}
\end{figure}

\noindent{\bf DLRover-RM Nears Well-Tuned Configurations.} To show the effectiveness of \system, we manually tuned the resource configurations for jobs without the support of  \system until they almost reached the best throughput. 
Fig.~\ref{fig:jct} shows that \system yields comparable JCTs to well-tuned settings. For instance, the JCT for model-X with \system is 27.74 minutes -- 1.4\% higher than the well-tuned counterpart. Note that manual tuning is a time-consuming \textit{trial-and-error} approach. For example, for Model-X, we re-run the job for more than 10 times to achieve its optimal configuration. The results demonstrate that \system's three-stage algorithm (\S\ref{subsec:3-stage_auto-scaling_optimization three-stage-algorithm}) can accurately capture/allocate the resource demands for different types of DLRM training jobs.
 
\noindent\textbf{DLRover-RM Beats Traditional Schedulers.}
Fig.~\ref{fig:jct} depicts that jobs under \system consistently achieve shorter JCT than ES and Optimus. On average, there is a 17.7\% or 28.5\% improvement compared with ES or Optimus. The results show that by considering the unique lookup operations and memory demands in DLRMs, \system can effectively schedule DLRM training jobs to improve the training efficiency with shorter JCT.

\begin{figure}[h]
\includegraphics[scale=0.125]{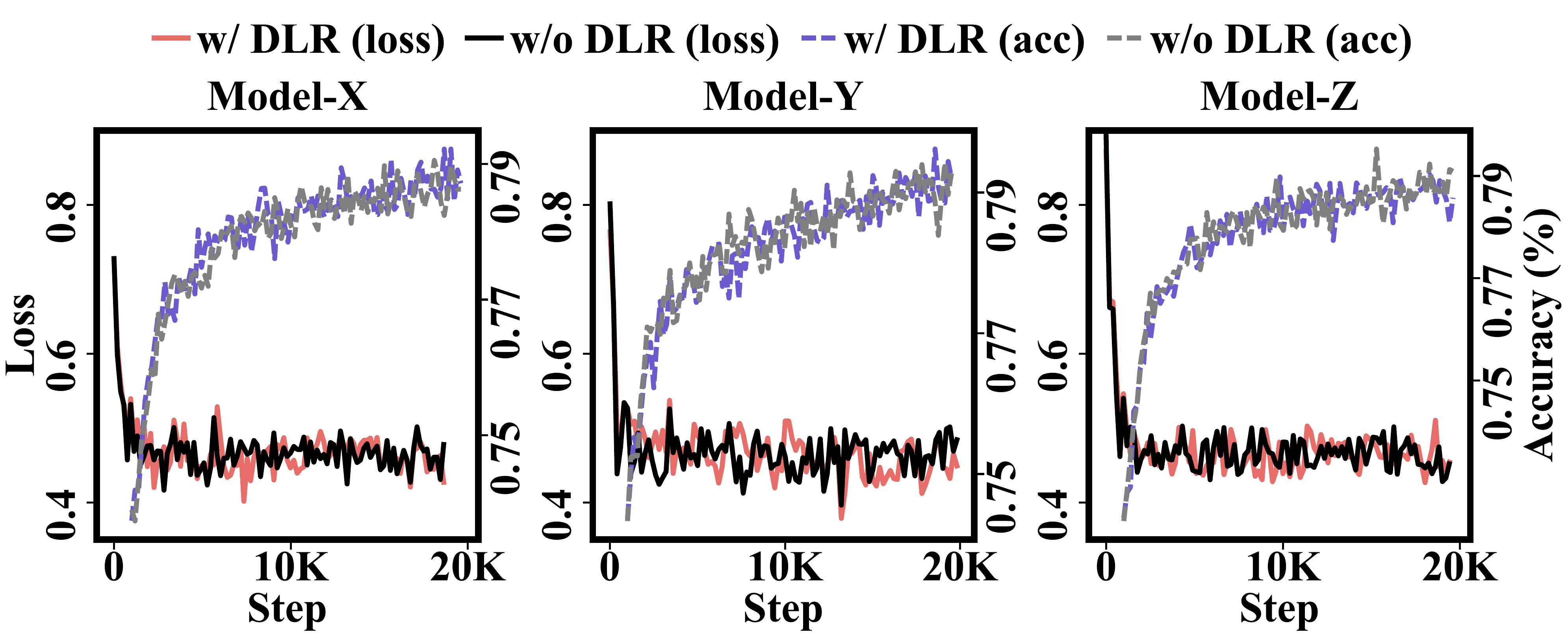}
\label{fig:train-loss}
\vspace{-2em}
\caption{DLRover-RM preserves the model performance (i.e., accuracy and loss) on different DLRM training jobs.}
\label{fig:loss}
\vspace{-2.5em}
\end{figure}

\noindent{\bf DLRover-RM Preserves Model Convergence.} In this experiment, we used 90\% of the Criteo dataset as the training set and the remaining as the test set. 
As shown in Fig.~\ref{fig:loss}, \system does \textit{not} compromise the model performance (i.e., convergence and accuracy) compared to jobs whose resource configurations are well-tuned. This verifies that the proposed \textit{dynamic data sharding} mechanism prevents inconsistent model quality caused by elastic operations (\S\ref{subsec: dynamic_data_sharding}).

In summary, \system enhances the training efficiency of various DLRM models by reducing JCT while maintaining model convergence. In the following section, we conduct an in-depth analysis of the separate effectiveness of the main components in \system.

\vspace{-2em}

%Combining the above experiments, 

\subsection{Ablation Study}
%In this section, we shows that each constituent part of the three-stage algorithm (\S \ref{subsec:3-stage_auto-scaling_optimization three-stage-algorithm}) in \system contributes significantly to its overall performance. 

\begin{figure}[h]
    \vspace{-1.5em}
    % \hspace{-1em}
      \subfigure[Worker] {
        \includegraphics[scale=0.125]{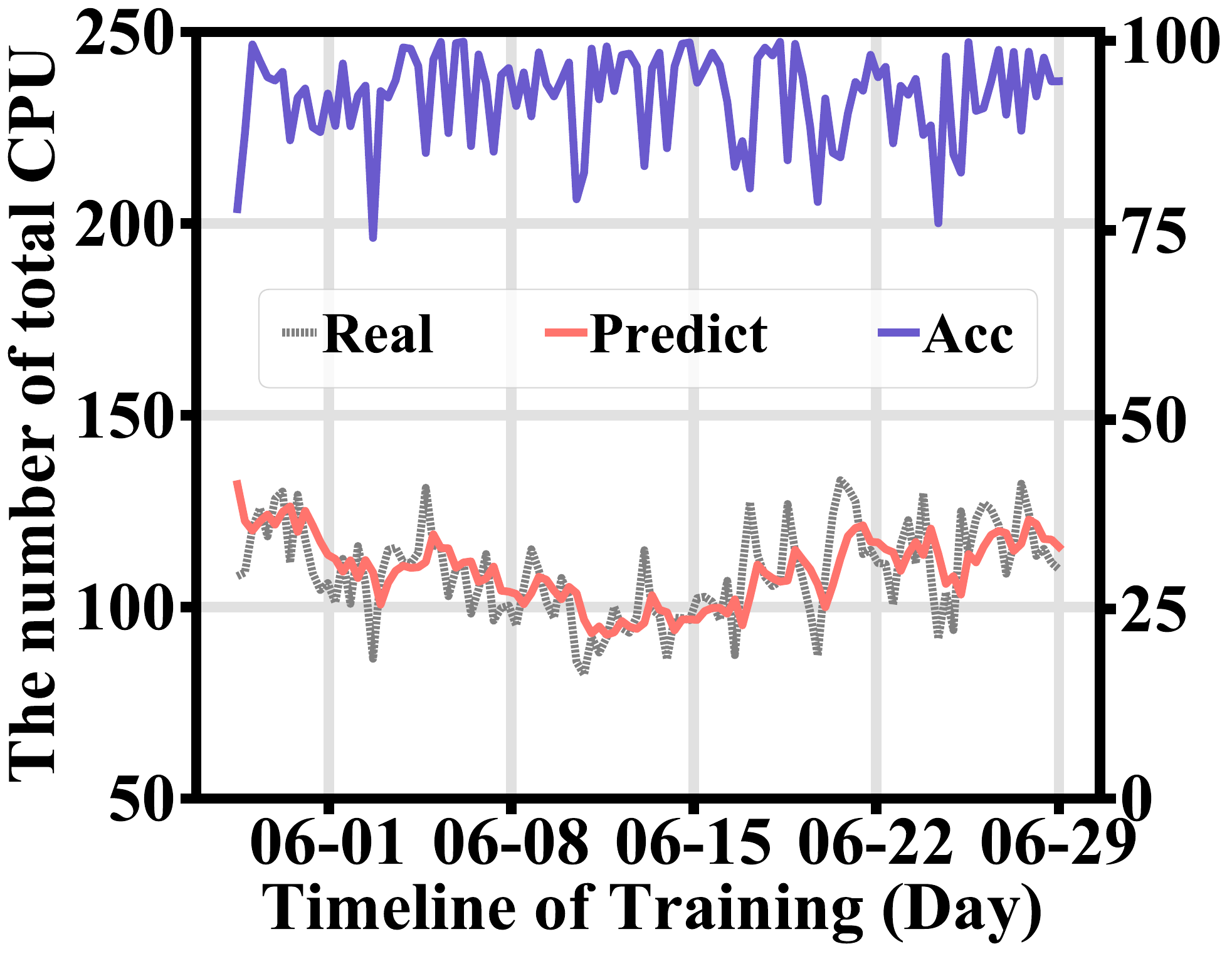}
        \label{fig:warmstart-userA}
      }
    \hspace{-1em}
      \subfigure[PS] {
        \includegraphics[scale=0.125]{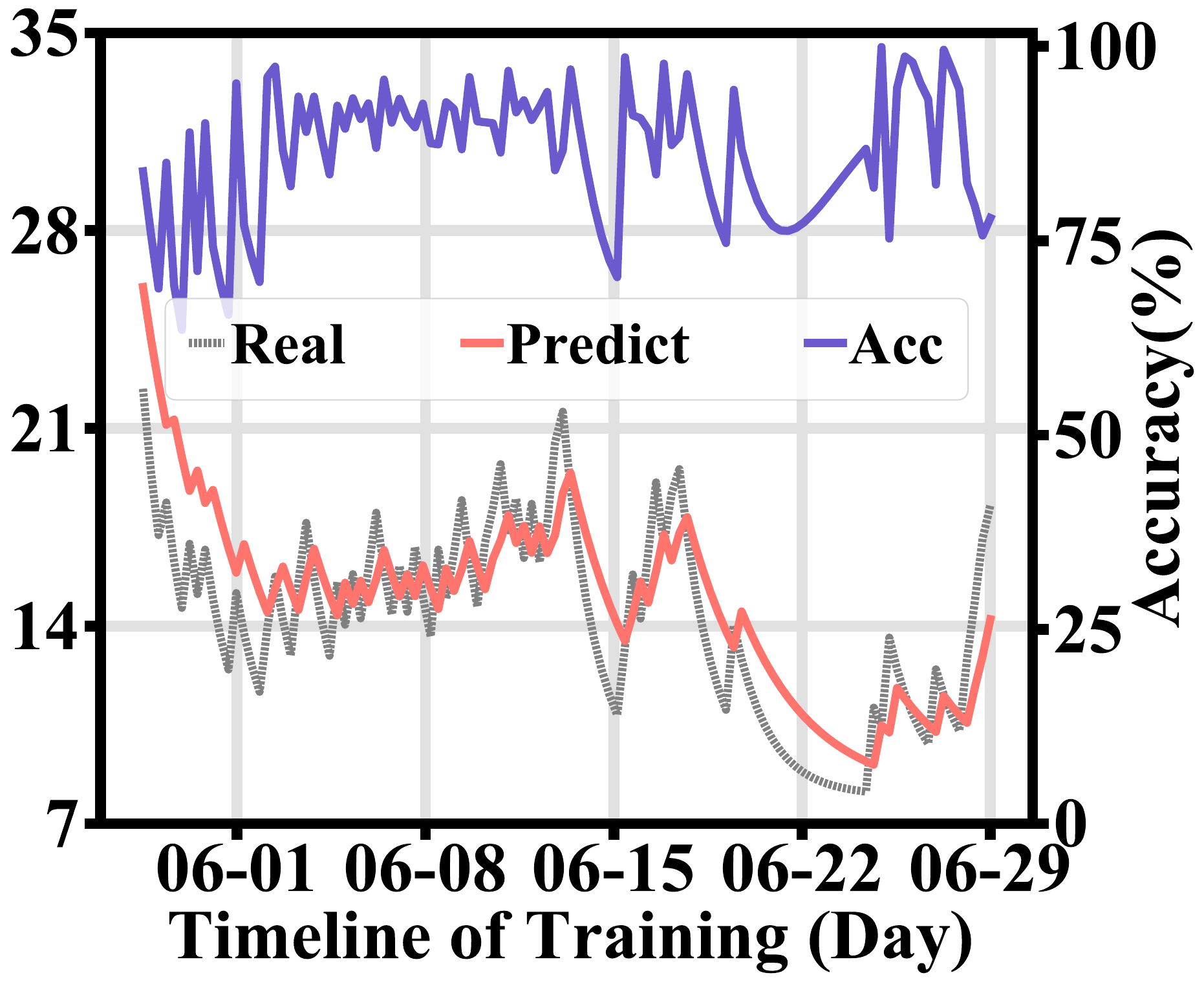}
        \label{fig:warmstart-userB}
      }
      \vspace{-0.2in}
      \caption{DLRover-RM's \textit{warm-starting} strategy provides a resource allocation close to the final
configuration.}
      \label{fig:warmstart}
    \vspace{-1.5em}
\end{figure}

\noindent{\bf Warm-starting.} To demonstrate the effectiveness of the \textit{warm-starting} algorithm (algorithm \ref{algo:data-driven-resource-setup}), we collected one month's job training data from a user within the production cluster. Fig.~\ref{fig:warmstart}  shows that \system, with the \textit{warm-starting} algorithm, provided initial resource allocation very close to the final configuration. On average, the accuracy of \system's initial and final configurations for workers and PSes are respectively 92\% and 85\%. It is because, based on similarity analysis, \system's \textit{warm-starting} algorithm extracts the most matching jobs from the users' historical task data to serve as guidance; 
%The benefit of this approach is that, 
With a good initial resource configuration, \system reduces the number of scaling. Thus, we reduce the time of scaling jobs from initial resource allocation to optimal resource allocation. Based on the statistics from the production cluster logs, compared to a cold-start approach (adjusting resources from zero), the scaling time was reduced by an average of 26\%. 

\begin{figure}[t]
% \vspace{-1em}
\includegraphics[scale=0.125]{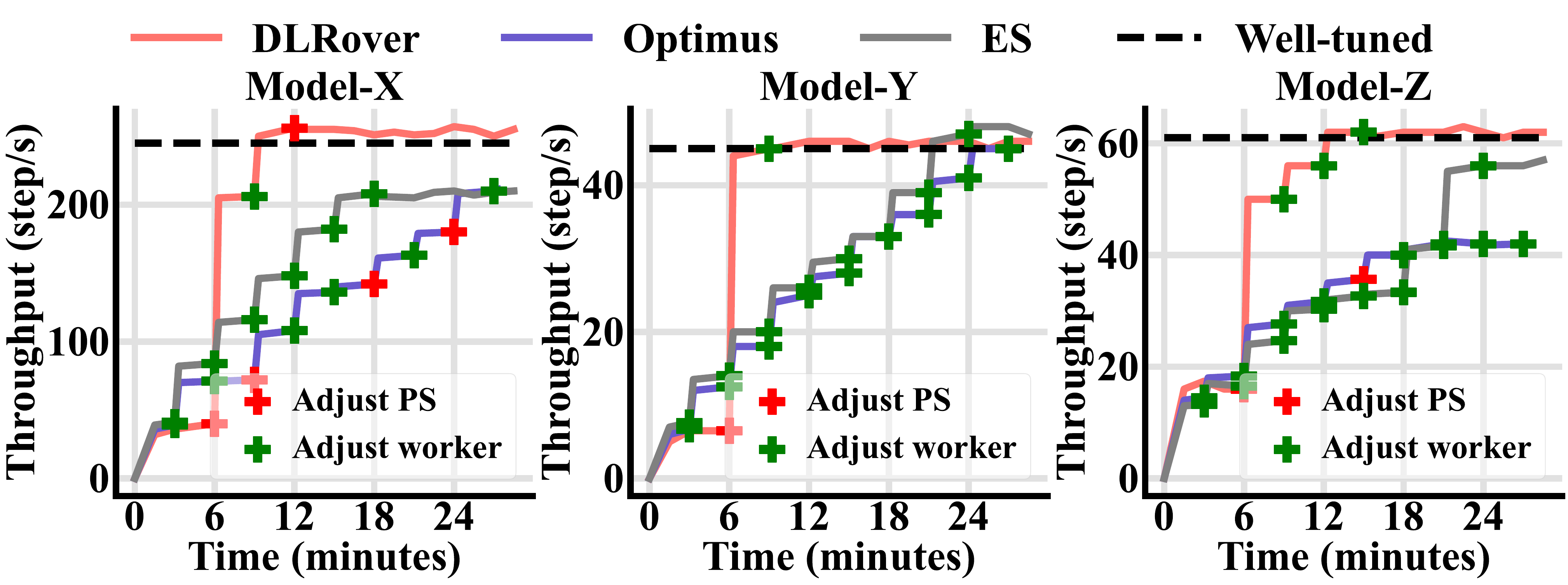}

\vspace{-1em}
\caption{DLRover-RM's scheduling algorithm achieves higher throughput in less time.}
\vspace{-1.5em}
\label{fig:train-throughput}
% \label{fig:throughput}
\end{figure}

\noindent{\bf Auto-Scaling.} To evaluate the effectiveness of auto-scaling strategies, we trained DLRMs from scratch (i.e., cold-started, thus removing the effect of \system's \textit{warm-starting}) with different schedulers. Every three minutes, schedulers adjusted the resources of PSes or workers based on the runtime information. Fig.~\ref{fig:train-throughput} shows that, compared to ES and Optimus, \system can achieve higher throughput within the same time period. For instance, for model-X, \system achieves a throughput of 250 steps/second after running for 12 minutes, while others still stay at the throughput of 100-150 steps/second. This is because \system considers unique lookup operations of DLRM training (in Eqn.~\ref{eq:emb}). %achieved better resource predictions. 

\begin{figure}[h]
\vspace{-1.5em}
\centering
\subfigure[validation on $w/p$]{
    \includegraphics[scale=0.1]{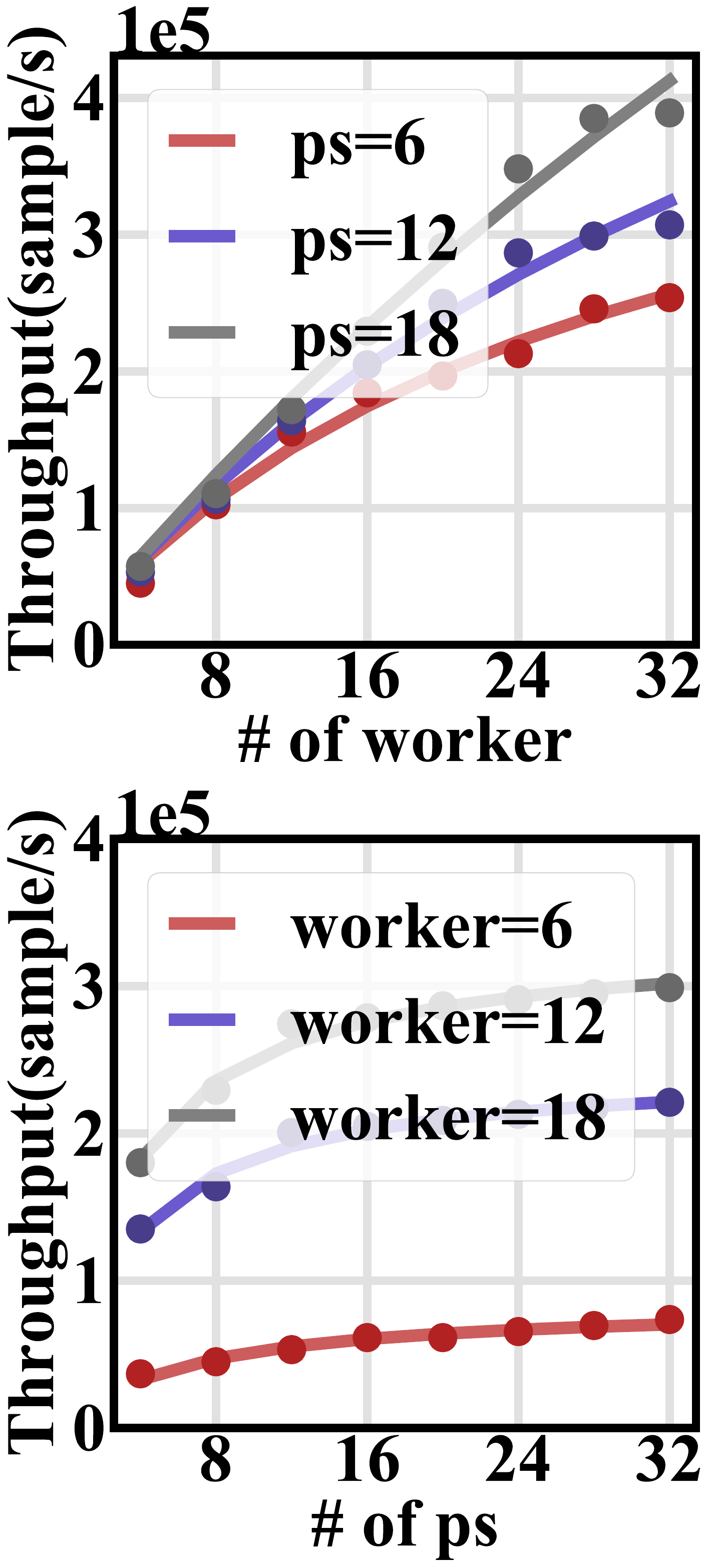}
}
\subfigure[validation on $\lambda_{w}$]{
    \includegraphics[scale=0.1]{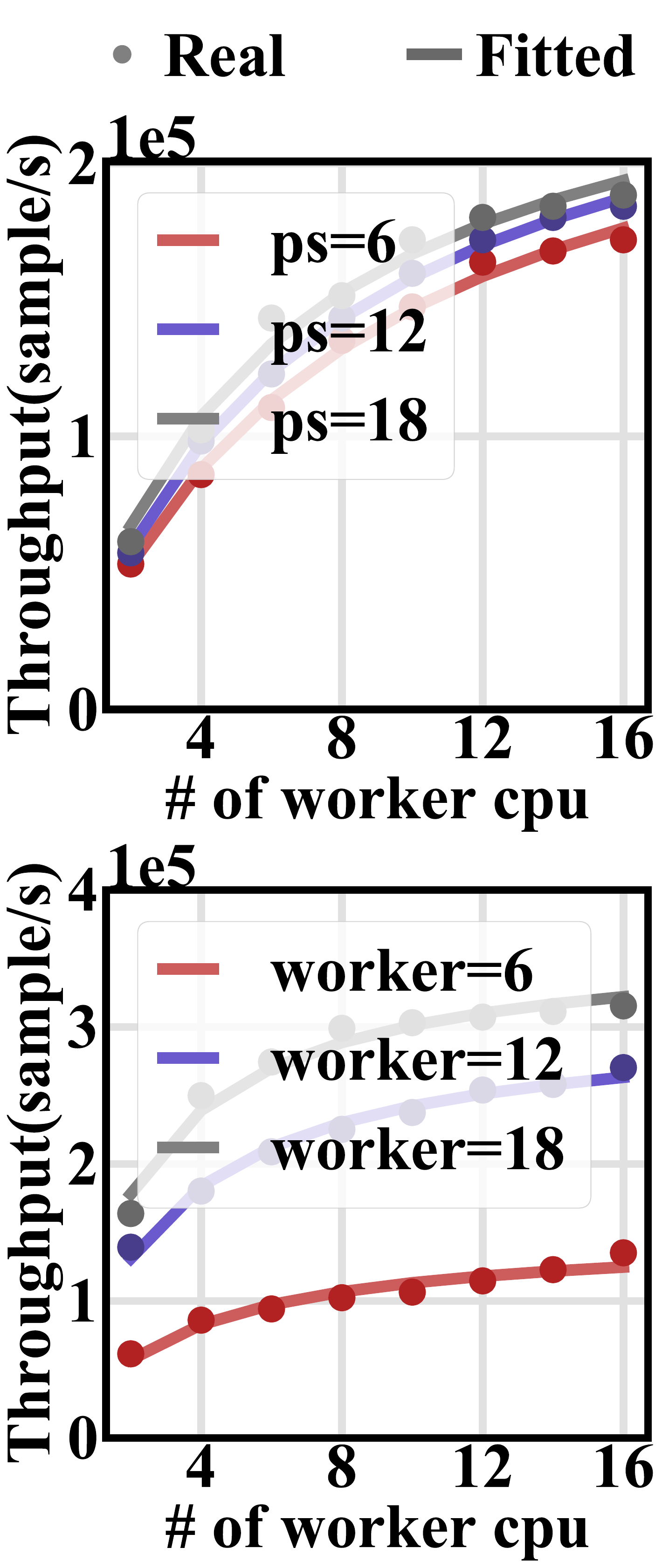}
}
\subfigure[validation on $\lambda_{p}$]{
    \includegraphics[scale=0.1]{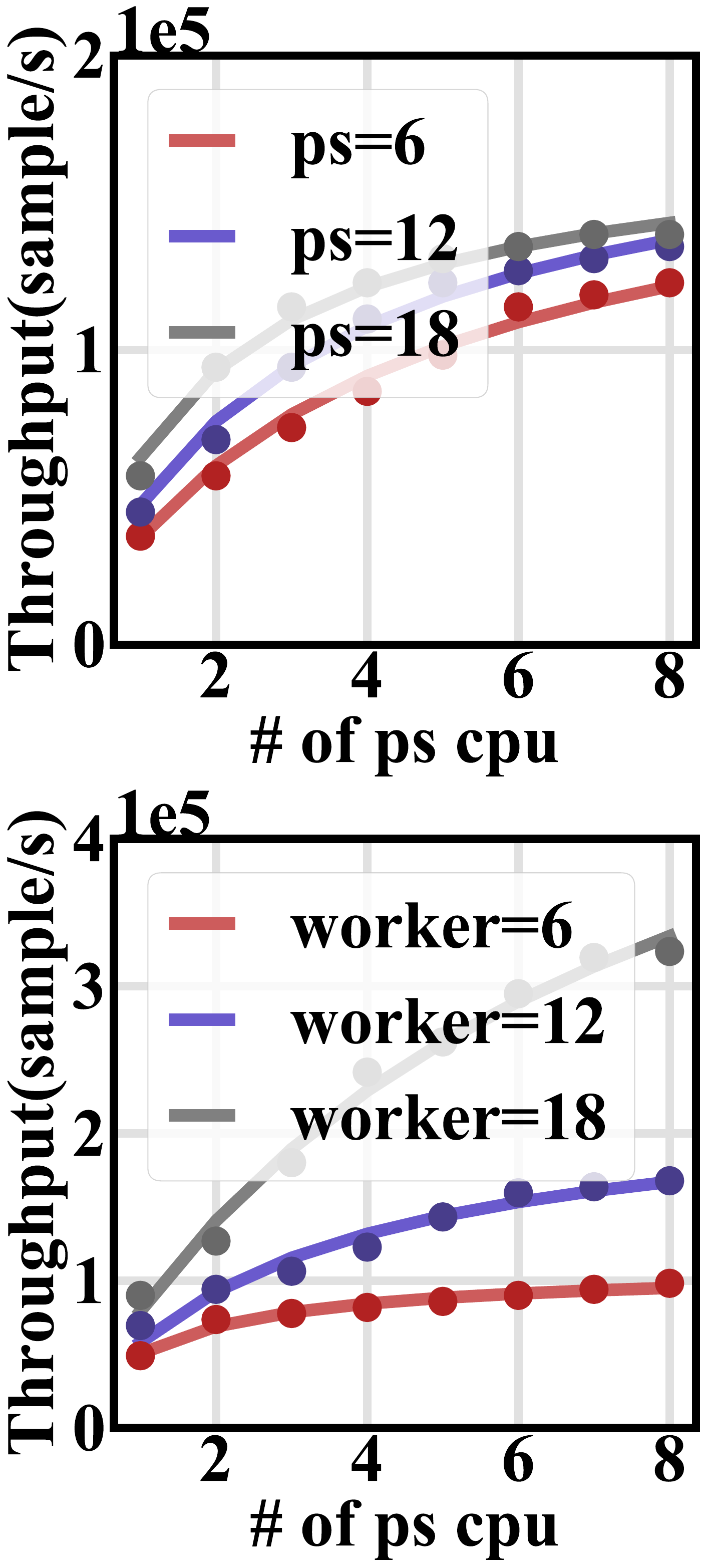}
}
\vspace{-1em}
\caption{Sampled data points and the fitted curves of the throughput prediction model. Under the setups with varying numbers of workers and PS, DLRover-RM accurately predicts the throughput when adjusting resource variables. }
\vspace{-1em}
\label{fig:model_fitting}
\end{figure}

To further verify this, we validate the proposed throughput prediction model (Eqn.~\ref{eq:throughput}). We sampled a set of training data points under different resource setups (different $(p,w,C_w,C_p )$). Then, we use NNLS \cite{NNLS} to find $\alpha$s~and~$\beta$s that best fit the collected data points.  As shown in Fig.~\ref{fig:model_fitting}, our model can closely describe the relationship between the training throughput and resource configurations. We report coefficients in  Eqn.~\ref{eq:throughput} as: $\alpha_{grad} = 3.48$, $\alpha_{upd} = 2.36$, $\alpha_{lookup} = 2.45$, $\alpha_{sync} = 0.68$, and 2.45 for the sum of $\beta$.

\begin{figure}[h] 
    \centering
    \hspace{-0.15cm}
    \includegraphics[width=\columnwidth]{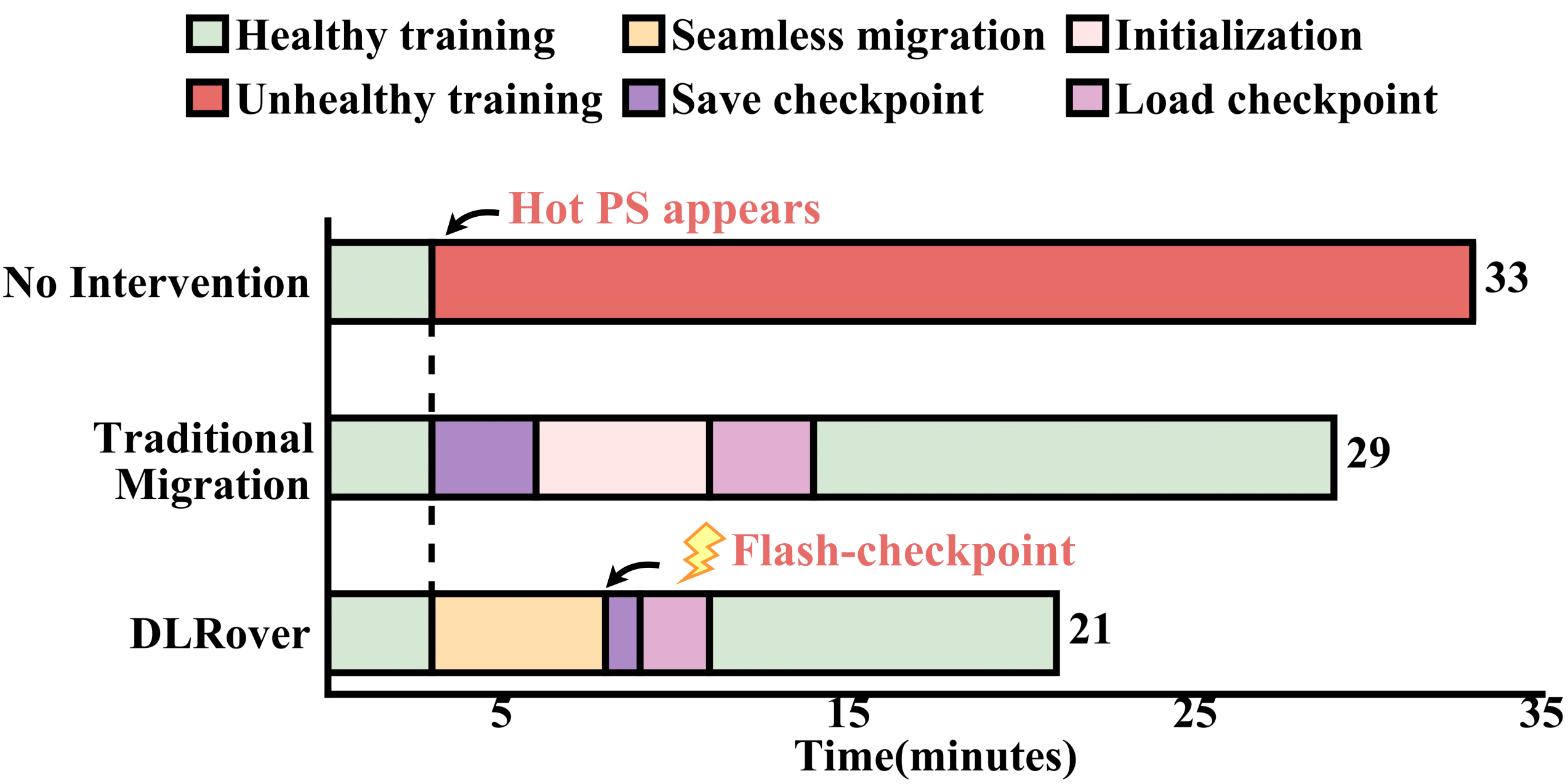}
    % \caption{The proportion of calculation time for DLRMs.} 
    \hspace{-0.15cm} 
    \vspace{-1em}
    \caption{When a hot PS occurs, three distinct strategies result in different JCT: The "no intervention" approach continues training under an unhealthy state. The "traditional migration" method employs a \textit{stop-and-restart} strategy, while \system utilizes the mechanism of \textit{seamless migration} and \textit{flash-checkpoint}, significantly reducing overhead (\S\ref{subsec:seamless_migration}). }
    \label{fig:hot_ps}
    \vspace{-1.5em}
\end{figure} 

\begin{figure}[h] 
    \centering
    \hspace{-0.15cm}
    \includegraphics[width=\columnwidth]{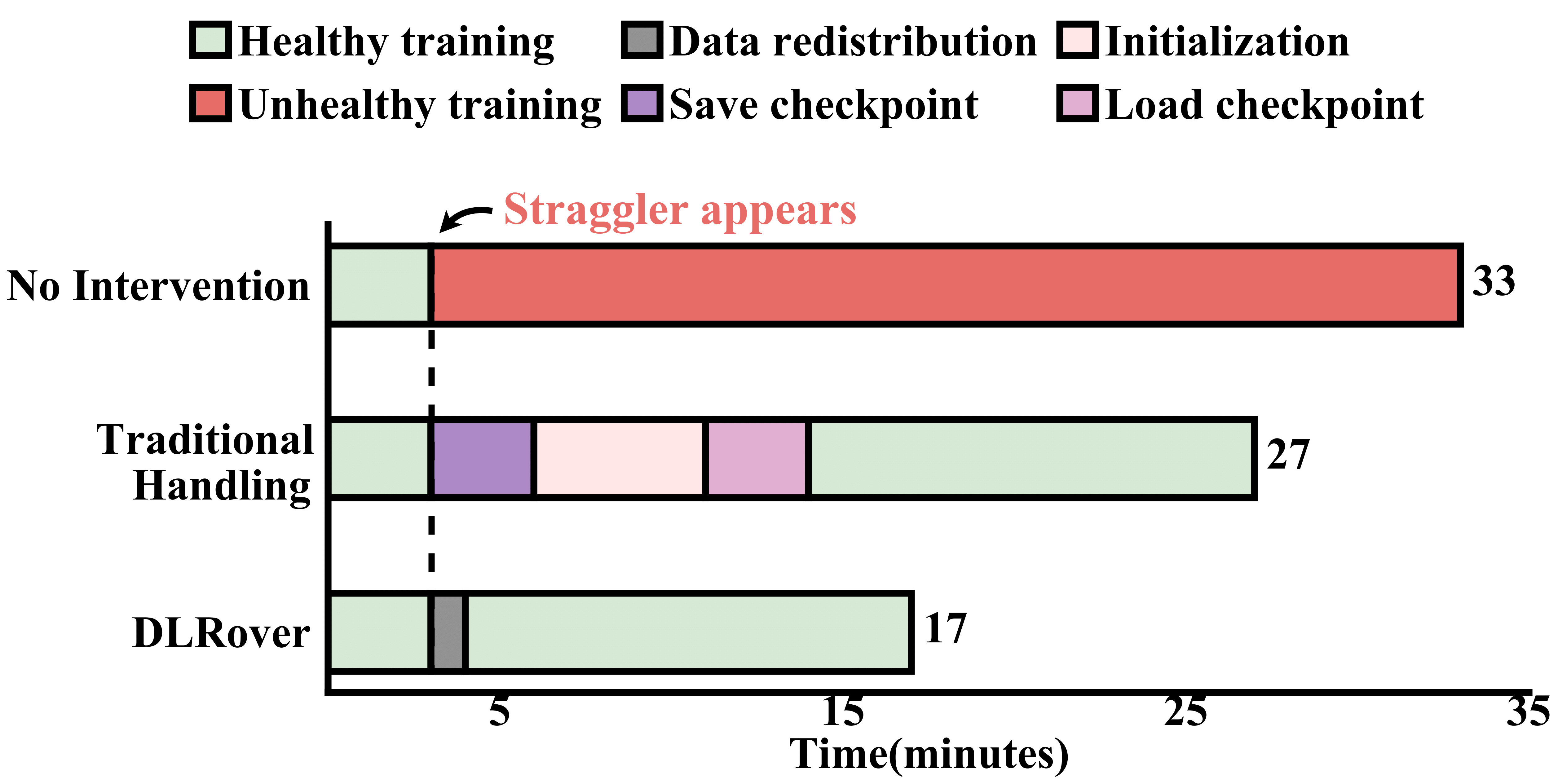}
    % \caption{The proportion of calculation time for DLRMs.} 
    \hspace{-0.15cm}
    \vspace{-1.5em}
    \caption{When a straggler occurs, three distinct strategies result in different JCT: The No Intervention approach persists in training under an unhealthy state. The Traditional Handling method employs a `stop-and-restart` strategy, while DLRover-RM redistributes the data shard for the straggler with the support of \textit{dynamic data sharding} mechanism (\S \ref{subsec: dynamic_data_sharding}), avoiding the restart of the job.} 
    % \Hao{For the straggler case, The no intervention and traditional handling method is as same as that of hot ps case. I have explained these in the caption of Fig. 11. I am not sure whether we still need to explain them again.}
    \label{fig:worker_straggler}
    \vspace{-2em}
\end{figure}

\noindent{\bf Instability Handling.} In this study, we verify the capability of \system to handle instability in the cloud (e.g., straggler and hot PSes). To simulate the worker/PS straggler cases, we randomly selected a worker or PS and set the CPU cores to 3\% of that in the well-tuned resource configuration.  

As shown in Fig.~\ref{fig:hot_ps},
for the hot PS  case, we observe that: 1) \system can reduce the JCT by 36.4\% and 27.6\% compared to the "no intervention" and "traditional migration" methods, respectively. 2) 
%Compared to 
Unlike "traditional migration", when stragglers are detected, \system enables continuous training (i.e., \textit{seamless migration}),
%This is achievable 
as \system restarts the job using an asynchronous approach that does not interrupt the job's training (\S \ref{subsec:seamless_migration}) -- saving about 5 minutes of the training time. 
3) With the \textit{flash-checkpoint} mechanism, \system saves 3 minutes in saving and loading checkpoints due to \system using shared memory to store checkpoints instead of communicating with RDS -- significantly reducing the communication overhead (\S \ref{subsec:seamless_migration}).
As shown in Fig.~\ref{fig:worker_straggler}, for the worker straggler case, we observe that 1) \system can shorten the JCT by 48.5\% and 37\% compared to the "no intervention" and "traditional handing" approaches, respectively. 2) Compared to "traditional handling", \system can also rapidly handle the straggler (within 1 minute) 
 and recover the healthy training instead of restarting the job. This is achieved by redistributing less data shard(s) to the straggler pod, with the support of dynamic data sharding mechanism (\S\ref{fig:dynamic_data_shard}).

\begin{figure*}[t]
\subfigure[Yearlong evolvement of CUR]{
    \includegraphics[scale=0.125]{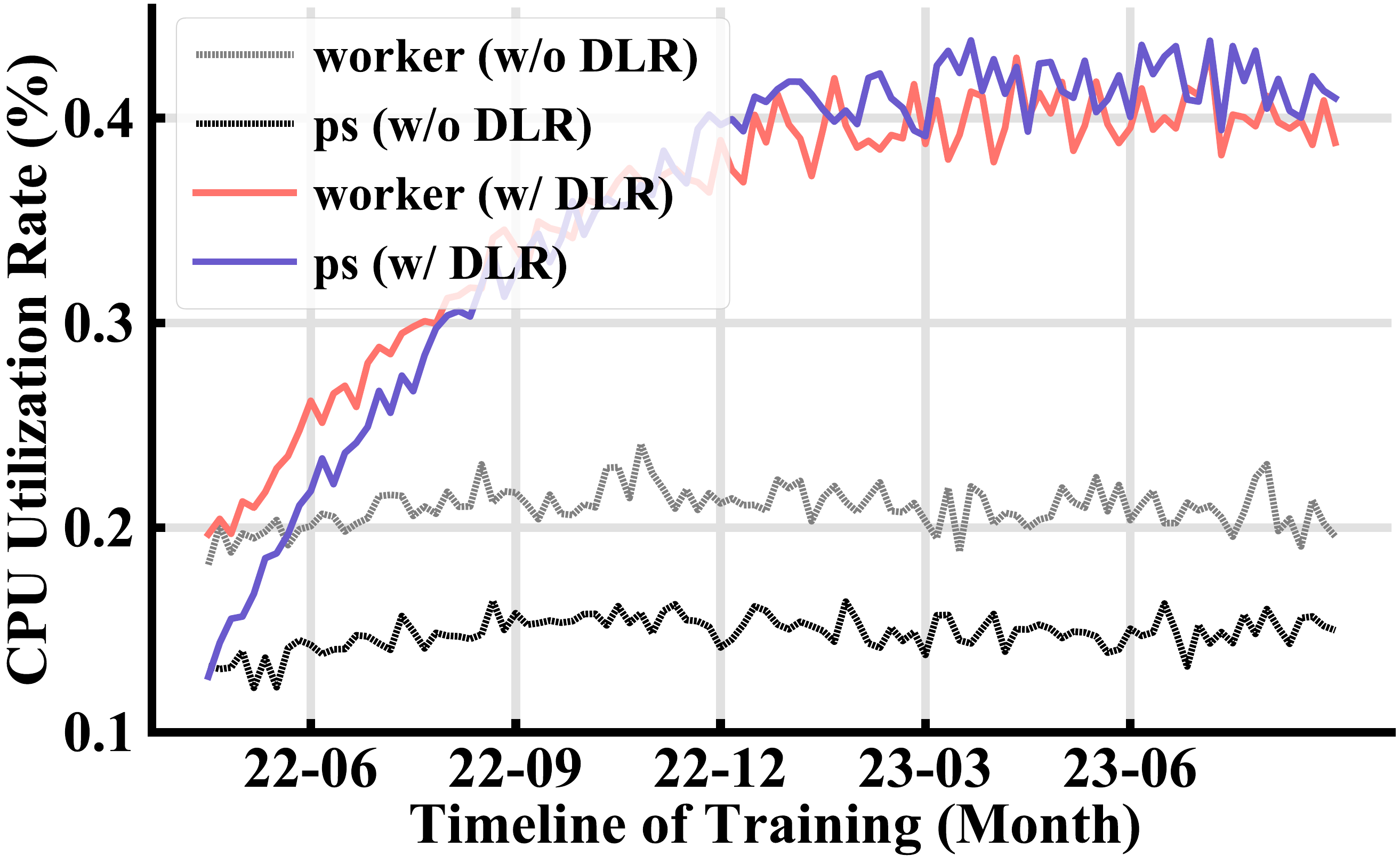}
    \label{fig:yearlong-cpu}
}
\subfigure[Yearlong evolvement of MUR]{
    \includegraphics[scale=0.125]{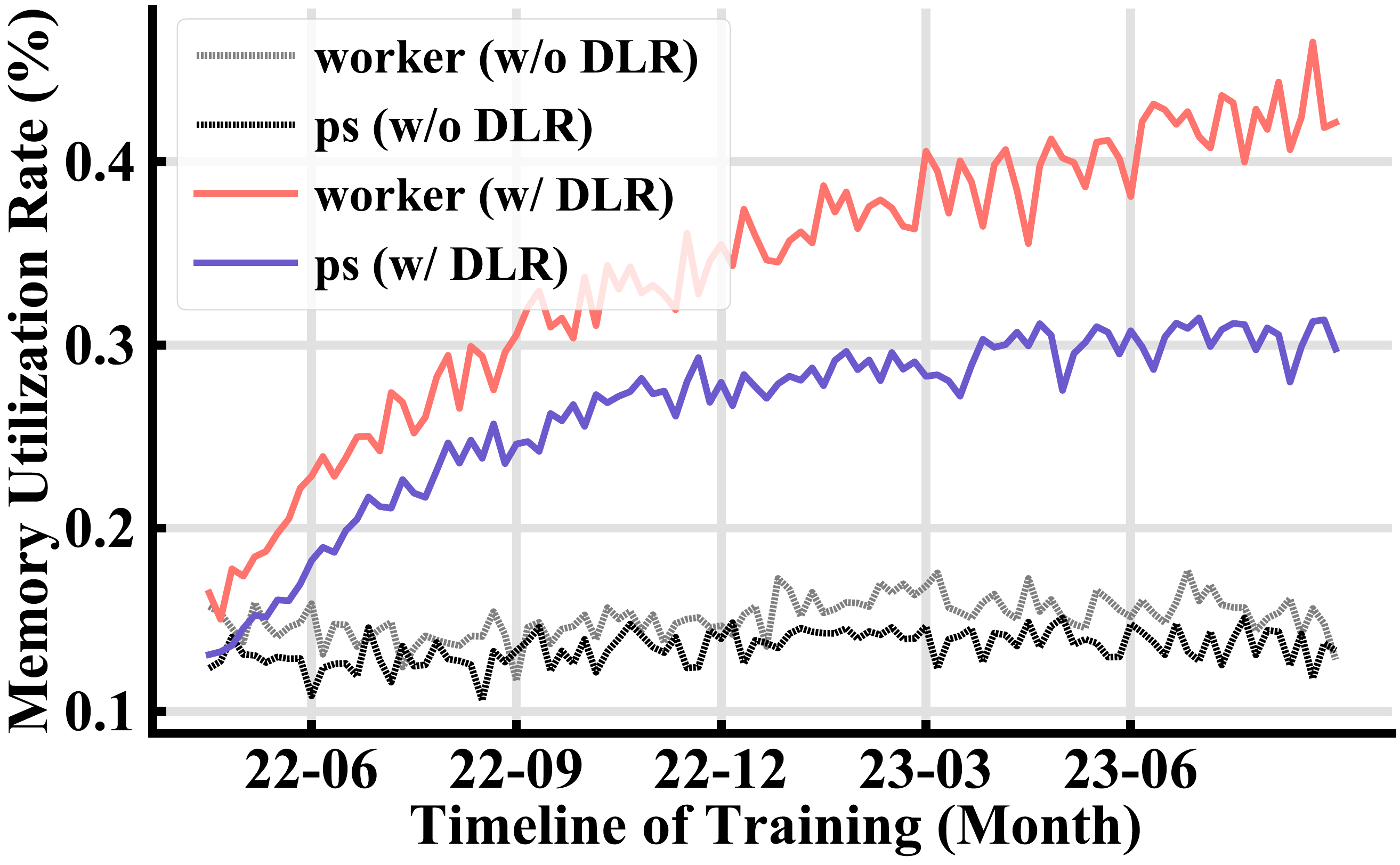}
    \label{fig:yearlong-memory}
}
\subfigure[Yearlong evolvement of JCR]{
    \includegraphics[scale=0.125]{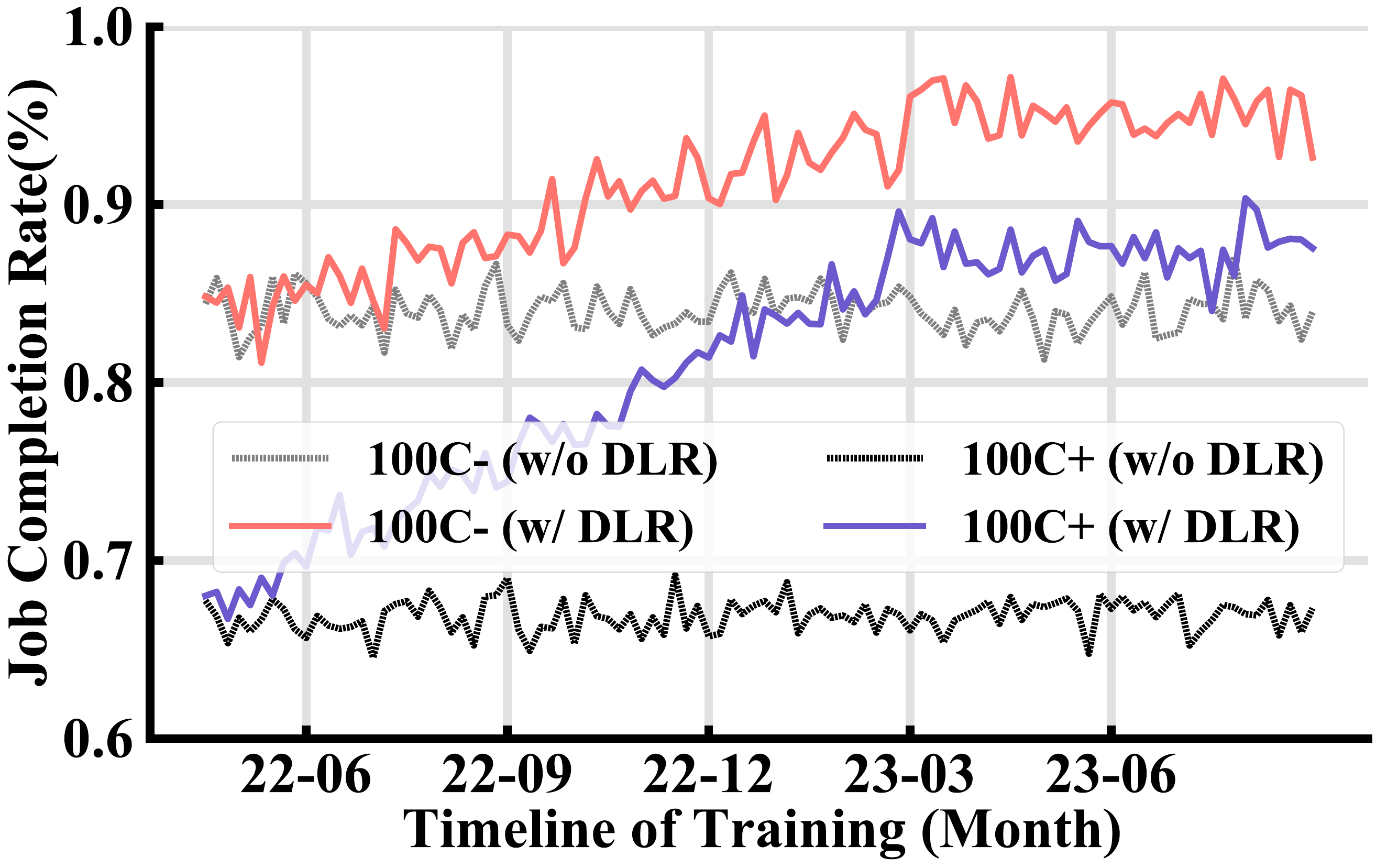}
    \label{fig:yearlong-jcr}
}
\vspace{-0.2in}
\caption{From  June 2022 to June 2023, 90\% jobs in the cloud-based cluster were progressively migrated to \system, leading to a significant increase in CPU utilization,  memory utilization, and Job completion rate. The black and gray lines represent jobs that cannot be migrated due to business reasons, which accounts for 5\% in the cluster.}
\vspace{-0.1in}
\label{fig:yearlong evalution}
\end{figure*}

% To assess the impact of various resource scheduling algorithms, such as \textit{ES} and \textit{Optimus}, we conducted an analysis of job throughput and Job Completion Time (JCT) results under similar setting. In the case of the \textit{ES} and \textit{Optimus} job schedulers, the adjustment interval used in all experiments was set to 3 minutes. This interval encompasses approximately 20 seconds for launching Pods, around 100 seconds for initializing training, and 60 seconds for sampling.

%at job level.

%  For the Wide \& Deep job, \system~finishes the adjustment within 12min and has better throughput than the other two algorithms. The throughput of Wide \& Deep job is limited by the number and CPU cores of PS and auto-scaling does not optimize PS at all. Optimus does add more PS but ignore the CPU cores of each PS. However there is a hot PS problem in Wide \& Deep job that can not be resolved with more PS. %The search time of Optimus is positive correlation to the adjustment number of parameter server/ worker. 
% Furthermore, \system~reduces the JCT by more than 30\% for Wide \& Deep and 10\% for xDeepFM (\autoref{fig:exp_jct_widedeep_jobs} and \autoref{fig:exp_jct_xdeepfm_jobs}).

% \begin{figure}[t] 
%     \centering
%     \includegraphics[scale=0.125]{figures/straggler_handling.pdf}
%     \caption{The impact of straggler on training} 
%     \hspace{-0.15cm}
%     \label{fig:straggler_handling}
%     \vspace{-1.5em}
% \end{figure} 

\vspace{-1em}
\subsection{Production Adoption and Evaluation}
\label{subsec:production_adoption_evaluation}
\system has been deployed in the cloud-based production cluster at \company since June 2022. Fig.~\ref{fig:yearlong evalution}
shows the change in CPU utilization rate, memory utilization rate, and job completion rate in the production cluster from June 2022 to June 2023, during which we have progressively migrated the jobs submitted with Kubeflow (i.e., a system without \system) to \system.
% \vspace{-3mm}

% \begin{center}
% \begin{table}[b]
%   \centering
%   \renewcommand{\arraystretch}{1.2}
% \caption{Profiling of DL jobs of \company}
%   \begin{tabular}{|c|p{1.8cm}|p{1.8cm}|}
%     \hline
%     \textbf{Job Statistic} & \textbf{TFJob} & \textbf{DLRover Auto}\\
%     \hline
%     Job Count & 5306 & 6109 \\
%     \hline
%     Avg steps  &268k & 249k \\
%     \hline
%     Avg parameters & 29.2 million & 34.5 million \\
%     \hline
%     Avg dataset size & 111 million & 103 million \\
%     % \hline
%     % Avg requested CPU  & 351 cores & 376 cores \\
%     % \hline
%     % Avg requested Memory & 596 GB & 564 GB \\
%     \hline

%   \end{tabular}
%   \vspace{-5mm}
% \label{table:cluster-profiling}
% \end{table}
% \end{center}

% DL training jobs have used about 100,000 CPUs and 100TB memory daily. 
\noindent{\bf Improved Resource Utilization.} As shown in Fig.~\ref{fig:yearlong-cpu} and \ref{fig:yearlong-memory}, as the jobs on the cluster were progressively migrated to \system, the CPU and memory utilization in the cluster exhibited a significant increase.  On average, the CPU utilization of workers and PSes increased from 19\% and 13\%  to 40\% and 41.4\%, respectively. In addition, the memory utilization of workers and PSes increased from 15.2\% and 13.8\% to 46.8\% and 31.1\%, respectively. 
%These increases significantly improve the DLRM training performance in the cloud. 

%Combining the above experiments, 
%We draw the conclusion that, 
%To summarize, after being deployed in our production environment, \system accelerates the job processing and achieves better resource utilization for DLRM training.
%at cluster-level.

\noindent{\bf Enhanced Fault Tolerance.}
{Fig.~\ref{fig:yearlong-jcr} shows that \system has significantly improved the job completion rate (JCR). Further, larger-scale training jobs can benefit more from \system: 1) The JCR for jobs needing fewer than 100 CPUs increased from 84\% to 95\% with \system; 2) the JCR for jobs that need more than 100 CPUs increased from 67\% to 87\% with \system}.

Moreover, we report the changes in slow training and failure rate within the cluster due to various reasons before and after using \system. As shown in Table~\ref{tab:failure_rate_comparison}, the proportion of job abnormalities (e.g., job failures and slow training) significantly decreases after using \system. These results show that \system, by employing the \textit{dynamic data sharding}, \textit{seamless migration}, \textit{flash-checkpoint}, and pre-adjustment-based \textit{OOM prevention} mechanisms, ensures robust fault tolerance for DLRMs training.

\begin{table}[h]
\centering
\vspace{-1em}
\caption{Failure rate comparison of before/after migration.}
\vspace{-1em}
\label{tab:failure_rate_comparison}
\begin{tabular}{|l|c|c|c|}
\hline
\textbf{Exceptions} & \multicolumn{1}{c|}{\textbf{Reasons}} & \multicolumn{1}{l|}{\textbf{w/o DLR}} & \multicolumn{1}{l|}{\textbf{w/ DLR}} \\ \hline
\multirow{2}{*}{Job Failure}   & OOM Errors & 4.7\% & 0.23\% \\ \cline{2-4} 
                               & Scheduling & 2\%   & 0.1\%   \\ \hline
\multirow{2}{*}{Slow Training} & Hot PSes   & 8\%   & 1\%     \\ \cline{2-4} 
                               & Worker Straggler & 7\%   & 0.7\%   \\ \hline
\end{tabular}
\vspace{-1em}
\end{table}

\begin{figure}[h]
    \vspace{-1em}
    \hspace{-1em}
      \subfigure[] {
        \includegraphics[scale=0.1]{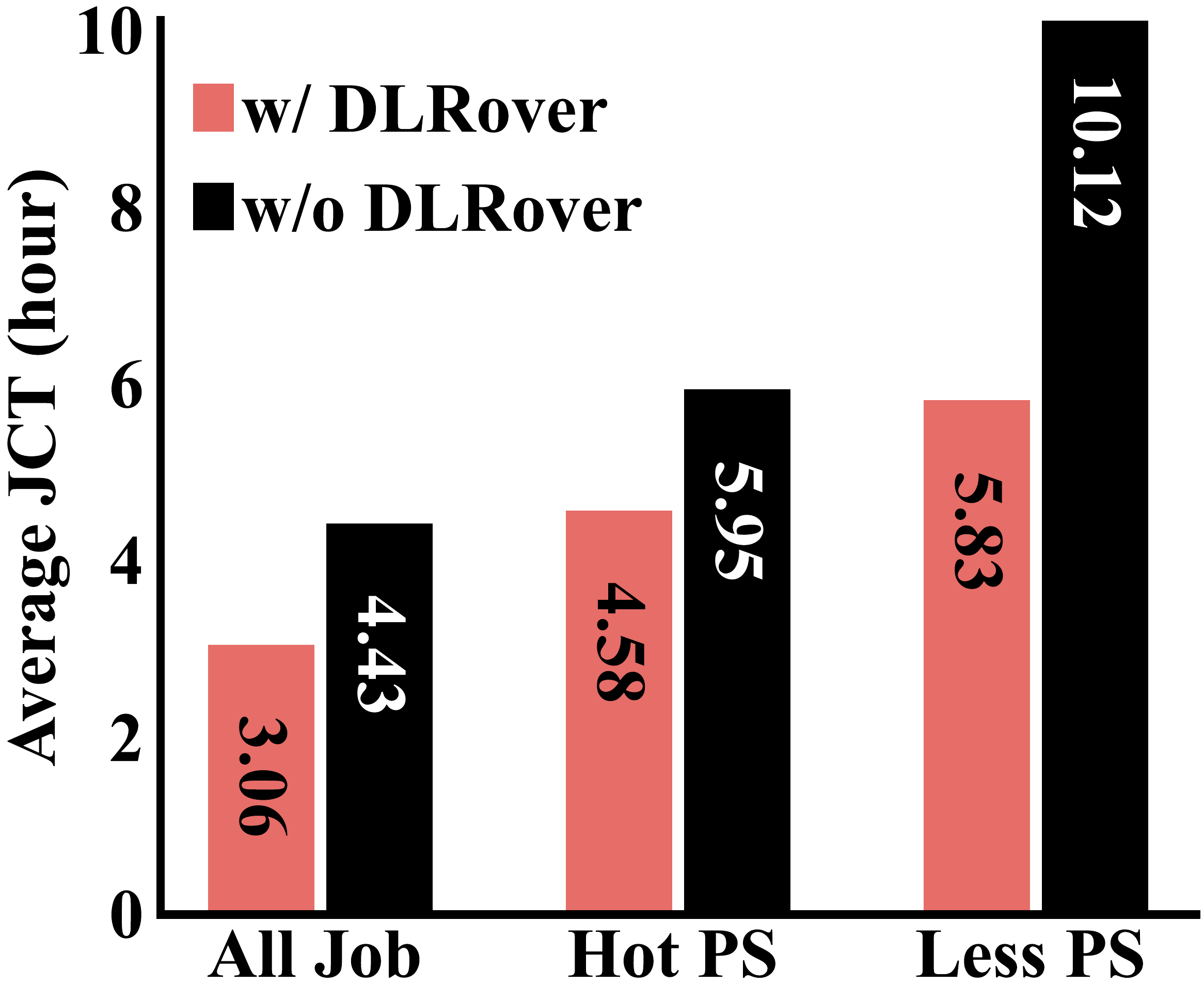}
        \label{fig:yearlong-jct-avg}
      }
    \hspace{-1em}
      \subfigure[] {
        \includegraphics[scale=0.1]{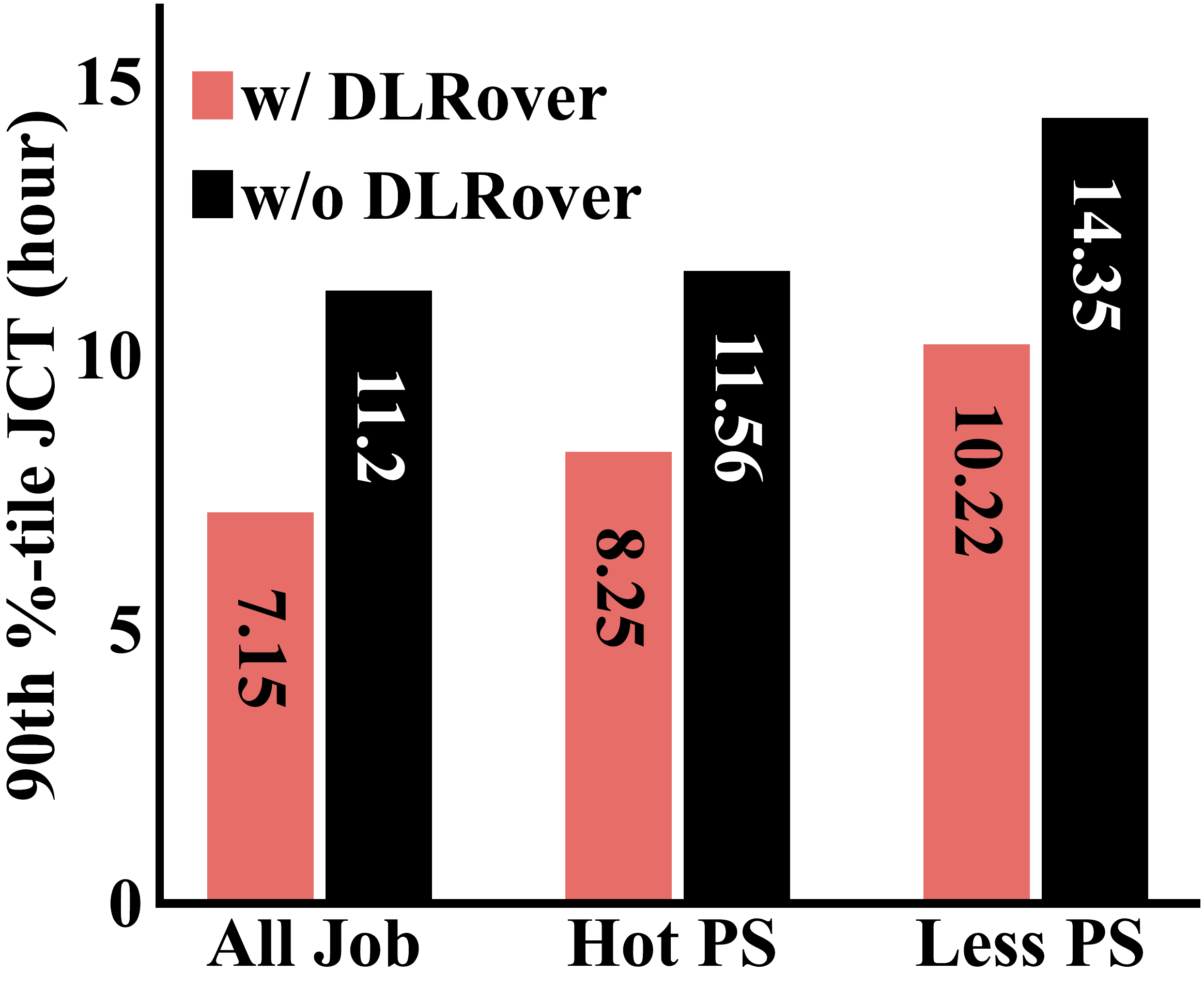}
        \label{fig:yearlong-jct-p90}
      }
      \vspace{-2em}
      \caption{DLRover-RM reduces cluster-level JCT.}
      \label{jct-yearlong}
      % \vspace{-0.4in}
      \vspace{-1.5em}
\end{figure}

\noindent{\bf Shortened JCT.} As \autoref{jct-yearlong} shows, \system~can shorten the JCT of DLRM training jobs significantly. We notice that 1) \system cut the median JCT and 90th\%-tile JCT for all jobs in the cluster by 31\% and 35.7\%, respectively. 2) For jobs with hot PSes, which occupy 13\% of all jobs, \system can shorten the median JCT and 90th\%-tile JCT of those jobs by 21\% and 28.6\%, respectively. 3) For jobs that are allocated with insufficient CPU for PSes (occupying 6\%), \system can shorten the median JCT and 90th\%-tile JCT of those jobs by 57\% and 28.7\%, respectively. 

\vspace{-0.5em}
\section{Related work}
\label{sec:related_work}
\textbf{Training of Deep Learning Recommendation.} Existing DLRM training systems focus on accelerating training speed and addressing memory pressure. For example, AIBox~\cite{zhao2019aibox} and HierPS~\cite{zhao2020distributed} overlap training execution on CPUs and GPUs, while using SSDs to store massive parameters of the model. Ekko~\cite{sima2022ekko} accelerates DLRM training over wide-area networks. Adnan et al.~\cite{memory_increase2} place highly accessed embeddings on GPU memory to reduce communication time. Gupta et al.~\cite{gupta2021training} compresses parameter gradients during model synchronization and squeezes activations and gradients across subnetworks during the forward and backward propagations. TTRec~\cite{yin2021tt} adopts tensor train decomposition to mitigate memory consumption. AdaEmbed~\cite{AdaEmbed} identifies the embedding rows with larger importance to improve model accuracy. 
AutoShard \cite{zha2022autoshard}, DreamShard \cite{zha2022dreamshard} and \cite{10.1145/3620678.3624661} seek the optimal embedding table sharding stragety to mitigate the lookup imbalances across devices. In contrast, the data sharding mechanism (\S\ref{subsec: dynamic_data_sharding}) in \system focuses on training data serving. FSDP\cite{pytorch} provides an industry-grade solution for large model data parallel training. In contrast, \system focuses on scheduling resources and provides robust fault tolerance for model training in a cloud environment, where failures are common and resource availability varies.  

\noindent\textbf{Automatic Resource Configuration.}
Automatic resource management is widely used in distributed data processing jobs on Hadoop~\cite{hadoop} and Spark~\cite{spark-nsdi}, and machine learning jobs on Spark MLlib~\cite{sparkmllib}. For example, Huang et al.~\cite{sparkml-auto} introduce an approach to configure the memory of large-scale ML on Spark automatically. Angel-PTM~\cite{nie2023angelptm} focuses on memory management optimization for large-scale transformer models. Eigen and fastflow~\cite{li2023eigen, um2023fastflow} explores resource efficiency optimization in large-scale public-cloud production environments.
%However, deep learning jobs are running customized algorithms with DNN layers, existing work is not suitable for DL training. 
Additionally, numerous endeavors have been dedicated to the automated configuration of the number of workers and parameter servers, such as Optimus, Pollux, and Tiresias \cite{Optimus, adaptdl, Tiresias}. Pollux~\cite{adaptdl} dynamically adjusts the number of workers and learning rate to improve the throughput for synchronous SGD. 
%However, they do not adjust the CPU and memory of each node.
%In the elasticity of 
Both Pollux and Tiresias must re-deploy all workers when adjusting resources, resulting in a long transition time. To minimize it, \cite{mlsys-elastic} starts new all-reduce operations only when new workers are ready and proposes a heuristic scaling to search the optimal number of workers.  For asynchronous training with the parameter server architecture, Optimus~\cite{Optimus} dynamically adds one worker or parameter server each time to maximize the cluster's performance without considering the transition time of elasticity. 
Regarding a large-scaling training job, the transition cost is not trivial because the number of workers and parameters is huge. 
\system is designed to conduct elasticity in a more effective way with little overhead. Furthermore, we plan to apply the LLM based sys optimization technique~\cite{gptunervldb} to improve the job initial configuration.

\noindent\textbf{Elastic Deep Learning Training.} There are deep learning frameworks that support elastic training. For instance, for asynchronous training, the PS training of TensorFlow~\cite{tensorflow2015-whitepaper} supports scaling workers at runtime. Using checkpoint, TensorFlow enables the elasticity for parameter servers. For synchronous training, there are Elastic DL~\cite{edl-on-gpu}, PyTorch-Elastic~\cite{torch-elastic}, and elastic Horovod~\cite{Horovod}. 
%To support elastic training, 
These systems typically restart the job by relaunching all pods~\cite{Optimus, adaptdl}. Users need to implement their dynamic data partition policy.
%When the number of workers changes, they must re-partition the whole dataset and restart an epoch~\cite{edl-on-gpu}. 
The re-partitioning may result in inconsistency of sample iterations if the dataset is huge. In contrast, \system has a \textit{dynamic data sharding} service and does not need to re-partition during elasticity. \system also ensures consistency when adjusting parameter servers by checkpointing unused data shards and model parameters leveraging memory storage like Gemini\cite{wang2023gemini}.

\noindent\textbf{Straggler Mitigation}. For a distributed asynchronous DLRM training job using parameter servers, the straggler could be a PS or worker because of hardware heterogeneity \cite{heter-cloud} or unbalanced data/parameter distribution \cite{tensorflow2015-whitepaper}. Existing works simply replace the slowest node with a new node to mitigate stragglers \cite{FlexRR, mlsys-elastic, Optimus}. However, this introduces additional overhead. Existing frameworks like Kubeflow \cite{kubeflow2023} can only set the same CPU and memory for the workers or PSes. In contrast, \system can adjust the workload and resources for various types of components on the fly based on updated elasticity decisions.

\vspace{-2em}

\section{Conclusion}
\label{sec:conclusion}
We have presented \system, a cloud-based DLRM training system. In designing \system, we have considered the unique characteristics of DLRMs and the practical challenges in a cloud environment. \system builds an accurate resource-performance model incorporating various runtime training information and develops a three-stage scheduling algorithm for elastic resource allocation and adjustment for DLRM training jobs. Moreover, \system offers a bunch of novel mechanisms to handle high cloud instability. 
%Its asynchronous checkpointing mechanism offers robust fault tolerance while minimizing performance impact. 
Our evaluation demonstrates the effectiveness of \system in reduced job completion time and increased resource utilization. 
%\system is publicly accessible and has been adopted by over ten major tech companies.
%to training jobs to improve training performance. Both the number of distributed nodes and the CPU/memory configuration of each node will be dynamically optimized during training jobs. After deploying \system~to k8s clusters, we have greatly improved job completion time, job completion rate, and cluster resource utilization.

%\clearpage
\normalem

\bibliographystyle{ACM-Reference-Format}
\bibliography{cites}

\end{document}